\documentclass[reprint, longbibliography,noeprint]{revtex4-1}

\usepackage[utf8]{inputenc}

\usepackage[lining,semibold]{libertine} 
\usepackage[libertine, cmintegrals, bigdelims, vvarbb]{newtxmath}

\usepackage{amsmath}
\usepackage{amsfonts}
\usepackage{mathrsfs}
\usepackage{gensymb}
\usepackage{bbm}
\usepackage{dsfont}

\usepackage{kbordermatrix}
\renewcommand{\kbldelim}{(}
\renewcommand{\kbrdelim}{)}

\usepackage{chemformula}
\usepackage[caption=false]{subfig}
\usepackage{chemfig}
\usepackage[version=3]{mhchem}

\usepackage{tikz}
\usetikzlibrary{matrix,positioning,decorations.pathreplacing}

\usepackage{soul}
\usepackage{xcolor}

\usepackage{scalerel} 

\definecolor{webgreen}{rgb}{0,.5,0}
\definecolor{webbrown}{rgb}{.6,0,0}
\definecolor{grigio}{rgb}{.85,.85,.85} 
\definecolor{RoyalBlue}{rgb}{0.0, 0.14, 0.4}
\definecolor{skyblue1}{rgb}{0.45,0.62,0.81}
\definecolor{skyblue2}{rgb}{0.2,0.39,0.64}
\definecolor{skyblue3}{rgb}{0.13,0.29,0.53}
\definecolor{scarlet1}{rgb}{0.93,0.16,0.16}
\definecolor{scarlet2}{rgb}{0.8,0,0}
\definecolor{scarlet3}{rgb}{0.64,0,0}

\definecolor{g}{gray}{0.50}

\usepackage{hyperref}
\hypersetup{%
    colorlinks=true, linktocpage=true, pdfstartpage=1, pdfstartview=FitV,%
    breaklinks=true, pdfpagemode=UseNone, pageanchor=true, pdfpagemode=UseOutlines,%
    plainpages=false, bookmarksnumbered, bookmarksopen=true, bookmarksopenlevel=1,%
    hypertexnames=true, pdfhighlight=/O,
    urlcolor=webbrown, linkcolor=RoyalBlue, citecolor=webgreen, 
    pdftitle={},%
    pdfauthor={Francesco Avanzini},%
    pdfsubject={},%
    pdfkeywords={},%
    pdfcreator={pdfLaTeX},%
    pdfproducer={LaTeX REVTeX}%
}

\newcommand{\norm}[1]{\lVert#1\rVert}

\newcommand{\chemspecies}{\alpha}

\newcommand{\rct}{\rho}

\newcommand{\setspecies}{Z}
\newcommand{\setspeciesa}{a}
\newcommand{\setspeciesb}{b}
\newcommand{\setgeneric}{v}
\newcommand{\setspeciesc}{\mathrm{ch}}
\newcommand{\indexx}{x}
\newcommand{\indexy}{y}

\newcommand{\setrct}{\mathcal R}
\newcommand{\setrcta}{A}
\newcommand{\setrctx}{X}
\newcommand{\setrcty}{Y}
\newcommand{\setrctcg}{\mathcal C}

\newcommand{\subx}{\mathcal X}
\newcommand{\suby}{\mathcal Y}

\newcommand{\conc}{\boldsymbol z}
\newcommand{\conca}{\boldsymbol a}
\newcommand{\concb}{\boldsymbol b}
\newcommand{\concc}{\boldsymbol c}
\newcommand{\concx}{\boldsymbol x}
\newcommand{\concy}{\boldsymbol y}
\newcommand{\concgeneric}{\boldsymbol v}

\newcommand{\curr}{\boldsymbol j}
\newcommand{\currs}{j}

\newcommand{\excurrs}{ I}
\newcommand{\curra}{\boldsymbol j^{\setrcta}}
\newcommand{\currx}{\boldsymbol j^{\setrctx}}
\newcommand{\curry}{\boldsymbol j^{\setrcty}}

\newcommand{\NA}{N_{\mathrm{A}}}

\newcommand{\matS}{\mathbb S}
\newcommand{\colS}{\mathbb S_\rct}

\newcommand{\matSa}{\mathbb S^{\setspeciesa}}
\newcommand{\matSb}{\mathbb S^{\setspeciesb}}

\newcommand{\matSaa}{\mathbb S^{\setspeciesa}_{\setrcta}}
\newcommand{\matSax}{\mathbb S^{\setspeciesa}_{\setrctx}}
\newcommand{\matSay}{\mathbb S^{\setspeciesa}_{\setrcty}}
\newcommand{\matSbx}{\mathbb S^{\setspeciesb}_{\setrctx}}
\newcommand{\matSby}{\mathbb S^{\setspeciesb}_{\setrcty}}

\newcommand{\mon}{\mathrm{M}}
\newcommand{\mona}{\mathrm{L}}

\newcommand{\A}{\mathrm{M}}
\newcommand{\E}{\mathrm{E_f}}
\newcommand{\Z}{\mathrm{Z_f}}
\newcommand{\CE}{\mathrm{E_c}}
\newcommand{\CZ}{\mathrm{Z_c}}

\newcommand{\matz}{\mathbb 0}

\newcommand{\chempotential}{\boldsymbol \mu}

\newcommand{\forces}{{F}}
\newcommand{\Aff}{\mathcal A}

\newcommand{\force}{\boldsymbol{\forces}}

\newcommand{\Sh}{S^{\text{H}}}
\newcommand{\Shannon}{H}
\newcommand{\Shb}{\Sh_\setspeciesb}
\newcommand{\Shx}{\Sh_\indexx}
\newcommand{\Shy}{\Sh_\indexy}

\newcommand{\Sz}{S_{\setspecies}}
\newcommand{\Sa}{S_{\setspeciesa}}
\newcommand{\Sb}{S_{\setspeciesb}}
\newcommand{\Sgeneric}{S_{\setgeneric}}
\newcommand{\Sch}{S_{\text{ch}}}
\newcommand{\Sph}{S_{\text{ph}}}

\newcommand{\epr}{\dot{\Sigma}}
\newcommand{\efr}{\dot{S}_e}
\newcommand{\epra}{\dot{\Sigma}^\setrcta}
\newcommand{\eprx}{\dot{\Sigma}^\setrctx}
\newcommand{\epry}{\dot{\Sigma}^\setrcty}
\newcommand{\efra}{\dot{S}_e^\setrcta}
\newcommand{\efrx}{\dot{S}_e^\setrctx}
\newcommand{\efry}{\dot{S}_e^\setrcty}

\newcommand{\pr}{p}
\newcommand{\nrm}{\norm{\concb}}
\newcommand{\nrmt}{\norm{\concb}}

\newcommand{\info}{\mathcal{I}_\setspeciesb}
\newcommand{\infoflowx}{\dot{\mathcal{I}}_\indexx}
\newcommand{\infoflowy}{\dot{\mathcal{I}}_\indexy}

\newcommand{\F}{\mathrm{F}}
\newcommand{\W}{\mathrm{W}}

\newcommand{\EF}{\dot{\mathcal{E}}}
\newcommand{\IF}{RT\nrmt\dot{\mathcal{I}}}

\newcommand{\dt}{\mathrm d_t}
\newcommand{\iindex}{i}

\makeatletter
\def\maketag@@@#1{\hbox{\m@th\normalfont\normalsize#1}}
\makeatother

\DeclareMathAlphabet{\mathpzc}{OT1}{pzc}{m}{it}

\begin{document}

\title{Information Thermodynamics for Deterministic Chemical Reaction Networks}
\newcommand\unilu{\affiliation{Complex Systems and Statistical Mechanics, Department of Physics and Materials Science, University of Luxembourg, L-1511 Luxembourg}}
\author{Emanuele Penocchio}
\email{emanuele.penocchio@uni.lu}
\unilu
\author{Francesco Avanzini}
\email{francesco.avanzini@uni.lu}
\unilu
\author{Massimiliano Esposito}
\email{massimiliano.esposito@uni.lu}
\unilu

\date{\today}

\begin{abstract}
Information thermodynamics relates the rate of change of mutual information between two interacting subsystems to their thermodynamics when the joined system is described by a bipartite stochastic dynamics satisfying local detailed balance.
Here, we expand the scope of information thermodynamics to deterministic bipartite chemical reaction networks, namely, composed of two coupled subnetworks sharing species, but not reactions.
We do so by introducing a meaningful notion of mutual information {\color{black}between different molecular features, that we express in terms of deterministic concentrations}.
This allows us to formulate separate second laws for each subnetwork, which account for their energy and information exchanges, in complete analogy with stochastic systems.
We then use our framework to investigate the working mechanisms of a model of chemically-driven self-assembly and an experimental light-driven bimolecular motor.
We show that both systems are constituted by two coupled subnetworks of chemical reactions.
One subnetwork is maintained out of equilibrium by external reservoirs (chemostats or light sources) and powers the other via energy and information flows.
In doing so, we clarify that the information flow is precisely the thermodynamic counterpart of an information ratchet mechanism only when no energy flow is involved.

\end{abstract}

\maketitle


\section{Introduction}

Engines commonly operate such that some components (e.g., pistons) directly interact with the power source to harvest energy, whereas some other components (e.g., wheels) produce the functionality the engine is designed for.
Chemical engines~\cite{amano2021chemeng}, such as molecular motors, are no exception~\cite{astumian1994,magnasco1994,bustamante2000}.
Indeed, they can be rationally described~\cite{bustamante2001,kolomeisky2007,lipowsky2008,mugnai2020,sivak2020} and designed~\cite{zerbetto2007,astumian2007,otto2015,amano2021chemeng,prins2021} as chemical reaction networks (CRNs) where energy-harvesting chemical~\cite{schliwa2003,boekhoven2010,stoddart2015,wilson2016,leigh2017pump,houdusse2020,amano2021pump}, photochemical~\cite{feringa1999,leigh2007,lehn2014,ragazzon2015,giuseppone2015,giuseppone2017,credi2021} or electrochemical~\cite{rapenne2009,stoddart2018,credi2021} processes are coupled with large-amplitude intramolecular motions or self-assembly reactions.
These CRNs are often bipartite~\cite{seifert2014bipartite,horowitz2014}, as the energy-harvesting processes act on some molecular properties (e.g., phosphorylation state, photo-isomerization state, oxidation/reduction state) while the processes realizing the engine's functionality act on different properties (e.g., position in space, assembly state).
When this is the case, the chemical species involved in the functioning of a chemical engine can be described with a double state ($\indexx, \indexy$), and the whole network can be split into two coupled subnetworks characterizing the interconversion of $\indexx$-states by one kind of processes (e.g., self assembly steps converting free states into assembled ones) possibly driven out-of-equilibrium by processes of another kind interconverting $y$-states (e.g., reaction with a chemical fuel acting as a substrate).

Assessing the thermodynamics of bipartite systems requires quantifying energy and information exchanges between the subnetworks by applying the tools of information thermodynamics~\cite{sagawa2010,seifert2014bipartite,horowitz2014,parrondo2015}.
Linear stochastic models of biochemical machines comprising unimolecular or pseudo-unimolecular reactions have been extensively studied from this perspective, revealing the fundamental role of information flows between different degrees of freedom in powering such single molecule machines~\cite{parrondo2013,seifert2014,loutchko2017,mugnai2021,ouldridge2017work,sivak2021infoflow}.
However, chemical engines may in general comprise nonlinear processes such as bimolecular reactions and are sometimes (almost always in case of synthetic ones) better described by deterministic dynamics expressed in terms of kinetic equations evolving experimentally measurable concentrations rather than probabilities.
At present, a theoretical framework able to systematically address the information thermodynamics of nonlinear deterministic CRNs is missing.

Very recently, we applied information thermodynamics to an autonomous synthetic molecular motor working at the deterministic level~\cite{amano2021info}.
Such analysis shows that information thermodynamics is in principle not limited to stochastic setups, but it strongly relies on the fact that the CRN describing the motor is only composed of unimolecular reactions.
Indeed, when this is the case, kinetic equations can be mapped into a Markov jump process on a linear network of states by normalizing the concentrations by the total concentration, which is a conserved quantity in linear networks.
Once this correspondence is in place, standard tools from information thermodynamics can be applied~\cite{horowitz2014}.
In particular, the mutual information~\cite{Cover2012}, a central quantity in information thermodynamics, can be defined at the level of normalized concentrations~\cite{amano2021info}.

In this paper, we go further by extending the framework of information thermodynamics to nonlinear deterministic CRNs, where multimolecular reactions cause the total concentration to vary in time.
We start by introducing the setup of deterministic CRNs in Section~\ref{sec:setup}, where the crucial notion of bipartite networks is formally defined and used to identify the two subnetworks, which influence each other despite the fact that each reaction pertains unambiguously to only one subnetwork.
In Section~\ref{sec:thermo}, thermodynamic quantities are introduced for both chemically-driven and light-driven CRNs, and the second law for nonequilibrium regimes is formulated.
The main result of this paper is obtained in Section~\ref{sec:info}, where the notion of mutual information for non-normalized concentration distributions is defined and used to formulate a second law for each subnetwork.
Crucially, these second laws show that the two subnetworks exchange free-energy trough information and energy flows.
Our result is analogous to the one obtained in the framework of stochastic thermodynamics of bipartite systems~\cite{seifert2014bipartite,horowitz2014}, but importantly information flow terms account here for the non-normalized concentration distributions.
Furthermore, the resulting expressions of the subnetworks' entropy production also consider the variation of the subnetworks entropy due to the standard molar entropy carried by the chemical species and the contribution of non-bipartite species which may be present.
To exemplify the use of our new framework, in Section~\ref{sec:applications} we apply it to two paradigmatic examples of out of equilibrium chemistry with nonlinear dynamics: a model for chemically-driven self-assembly~\cite{ragazzon2018,prins2021} and a light-driven bimolecular motor~\cite{ragazzon2015,sabatino2019,credi2021,asnicar2022,corra2021}.
We briefly recap the basis of their functioning while focusing on the insights brought by the information thermodynamic analysis.
Using numerical simulations, we also comment on their efficiency and we quantitatively investigate the correspondence between the concept of Brownian information ratchet~\cite{zerbetto2007,astumian2019} and (thermodynamic) information flows.

Our approach can be applied to any CRN with a bipartite structure and coupled to any kind of reservoir.
We restricted the presentation to the case of autonomous and homogeneous ideal dilute solutions, but extensions to nonautonomous~\cite{forastiere2020}, non-homogeneous~\cite{Falasco2018a,Avanzini2019a} and nonideal~\cite{Avanzini2021} CRNs are possible, as well as to cases where external species are continuously injected~\cite{avanzini2022flux}.
Based on the relevance of information thermodynamics as a tool to elucidate free-energy processing in the stochastic realm, we envision that our new framework will bring significant insights in the understanding of deterministic chemical processes, from synthetic molecular machines to complex biochemical networks.

{\color{black} 
Finally, we note that many forms of information processing are possible in CRNs~\cite{xiang2018,grozinger2019}: e.g., logic gates~\cite{green2017,wang2020implementing}, machine learning~\cite{qian2011,deutsch2018,ouldridge2022}, sensing~\cite{malaguti2021}, copying and proofreading~\cite{gaspard2008,pigolotti2015,ouldridge2017comp,ouldridge2019}, memories~\cite{padirac2012,boekhoven2022}. At this stage, how the present framework may be suitable to study all of them is left for future inquiries.}


\section{Setup}
\label{sec:setup}

CRNs are treated here as ideal dilute solutions composed of chemical species, identified by the label $\chemspecies\in\setspecies$, which undergo elementary~\cite{Svehla1993} or coarse-grained~\cite{Wachtel2018, Avanzini2020b, penocchio2021photo} chemical reactions, identified by the index $\rct\in\setrct$:
\begin{equation}\color{black}
\boldsymbol \chemspecies\cdot \boldsymbol \nu_{+\rct} \ch{->[ ][ $\rct$ ]} \boldsymbol \chemspecies\cdot \boldsymbol \nu_{-\rct}\,,
\label{eq_elementary_reaction}
\end{equation}
with $\boldsymbol \chemspecies =(\dots,\chemspecies,\dots)^\intercal$ {\color{black}denoting} the vector of chemical species{\color{black}, while $\boldsymbol \nu_{+\rct}$ and $\boldsymbol \nu_{-\rct}$ denote} the vectors of stoichiometric coefficients {\color{black} of reagents and products of reaction $\rct$.}
{\color{black} Throughout this paper, we will use single arrows like in Eq.~\eqref{eq_elementary_reaction} to denote reactions that are actually reversible: for every reaction $\rct\in\setrct$, the forward (resp. backward) reaction $+\rct$ (resp. $-\rct$) interconverts $\boldsymbol\chemspecies\cdot \boldsymbol \nu_{+\rct}$ (resp. $\boldsymbol\chemspecies\cdot \boldsymbol \nu_{-\rct}$) into $\boldsymbol\chemspecies\cdot \boldsymbol \nu_{-\rct}$ (resp. $\boldsymbol\chemspecies\cdot \boldsymbol \nu_{+\rct}$).
This choice will simplify the hypergraph representation of the two CRNs examined in Sec.~\ref{sec:applications} (see Figs.~\ref{fig:drivenSA} and~\ref{fig:photomotor}).}

In deterministic CRNs, the abundance of the chemical species is specified by the concentration vector $\conc=(\dots, [\chemspecies],\dots)^\intercal$, 
which follows  the rate equation
\begin{equation}
\dt \conc=\matS \curr\,,
\label{eq_dynamics_crns}
\end{equation}
where we introduced the stoichiometric matrix $\matS$ and the current vector $ \curr$.
Each $\rct$ column ${\colS}$ of the stoichiometric matrix $\matS$ specifies the net variation of the number of molecules for each species undergoing the $\rct$ reaction~\eqref{eq_elementary_reaction}, ${\colS} =  \boldsymbol\nu_{-\rct} - \boldsymbol\nu_{+\rct} $.
The current vector $ \curr=(\dots,\currs^\rct,\dots)^\intercal$ specifies the net reaction current for every $\rct$ reaction~\eqref{eq_elementary_reaction} as the difference between the forward and backward flux, i.e., $\currs^\rct = \currs^{+\rct} - \currs^{-\rct}$.
In \textit{closed} CRNs, where the chemical reactions involve only the chemical species $\setspecies$, the fluxes $\currs^{\pm\rct}$ depend only on the concentrations $\conc$.
For elementary reactions satisfying mass action kinetics~\cite{Groot1984,Laidler1987,Pekar2005}, $ \currs^{\pm\rct} = k_{\pm\rct}\conc^{\boldsymbol\nu_{\pm\rct}}$, with $k_{\pm\rct}$ the kinetic constants {\color{black}(hereafter, for every vectors ${\boldsymbol v}$ and  ${\boldsymbol w}$,  ${\boldsymbol v}^{\boldsymbol w}= \prod_i v_i^{w_i}$)}.
In \textit{open} CRNs, some chemical reactions exchange for instance matter and/or photons with external reservoirs. 
Thus, the corresponding fluxes $\currs^{\pm\rct}$ depend on the concentrations $\conc$ as well as on the coupling mechanisms with the reservoirs~\cite{avanzini2022flux}.
Note that, throughout the manuscript, we will omit for compactness of notation the functions' variables, e.g., $\conc = \conc(t)$ and $\currs^{\pm\rct} = \currs^{\pm\rct}(\conc)$ (if reaction $\rct$ is not coupled to any reservoir). 

\begin{figure}[t]
        \includegraphics[width=.49\textwidth]{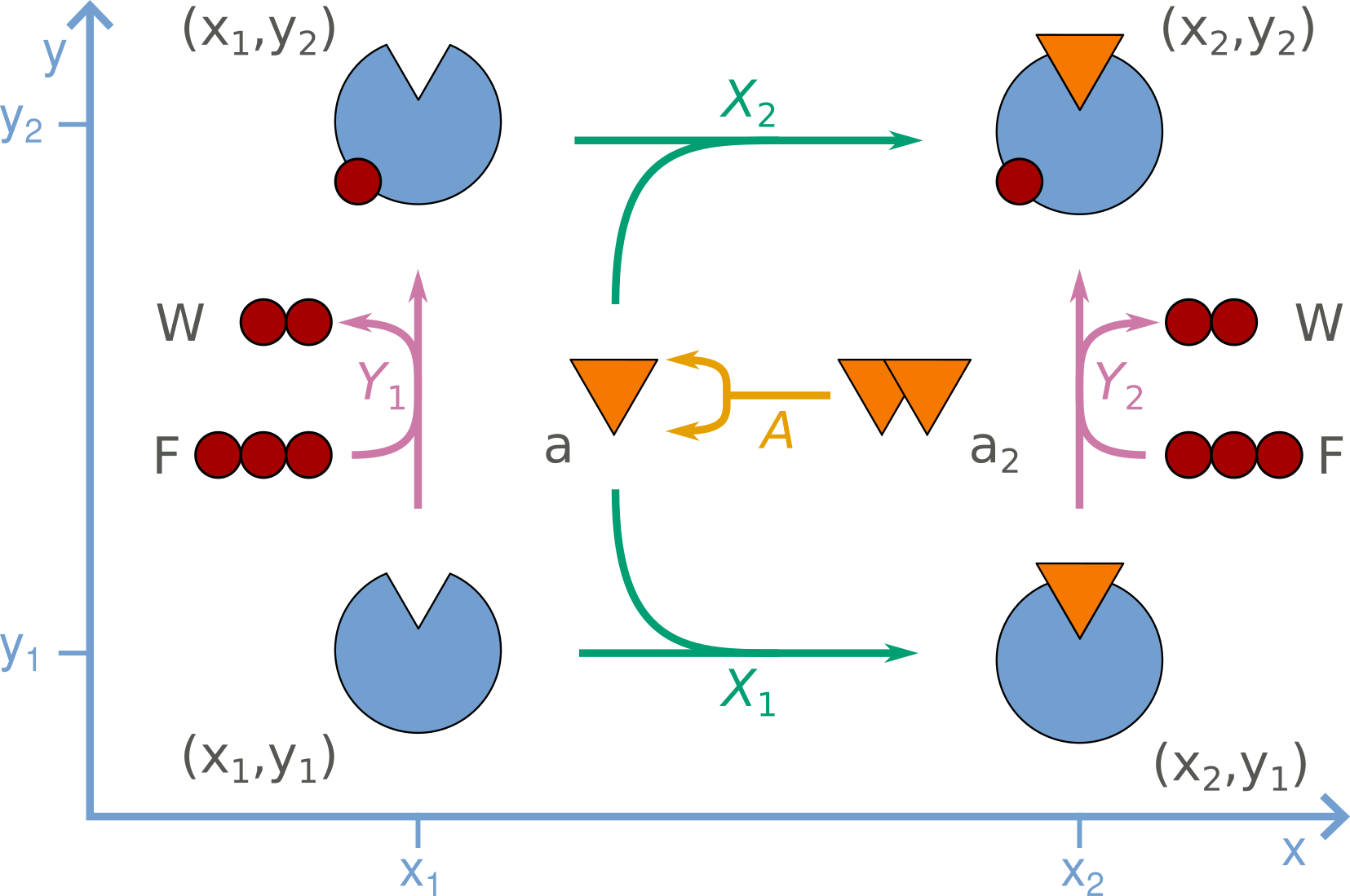}
        \centering
        \caption{\color{black}
        \textbf{An example of bipartite network.}
To illustrate our notion of bipartite structure in a CRN, we consider a generic molecule (blue ``Pac-Man'') able to bind two substrates.
One substrate (orange triangle, denoted $\ch{a}$) is generated from a precursor (denoted $\ch{a2}$) via reaction $A$: $\ch{a2} \rightarrow 2\ch{a}$ (notice the use of hypergraph notation where the production of two molecules of $\ch{a}$ is represented by a bifurcating arrow).
The substrate $\ch{a}$ can then bind the molecule via reactions $\setrctx_1$ or $\setrctx_2$, which both change its $\indexx$ state from $\indexx_1$ (free) to $\indexx_2$ (bound).
The other substrate (red sphere) is exchanged between the molecule and compounds $\F$ and $\W$ through reactions $\setrcty_1$ and $\setrcty_2$, which both change the $\indexy$ state of the molecule from $\indexy_1$ (free) to $\indexy_2$ (bound).
As a result, the molecule can be found in four double states $(\indexx,\indexy)$: $(\indexx_1,\indexy_1)$, $(\indexx_2,\indexy_1)$, $(\indexx_2,\indexy_2)$, $(\indexx_1,\indexy_2)$, which form the set of bipartite species (set $\setspeciesb$).
All other species in the network ($\ch{a2}$, $\ch{a}$, $\F$ and $\W$) can be treated either as ancillary species (set $\setspeciesa$) or, if their concentration is controlled by external reservoirs, as chemostatted species (i.e., controlled parameters).
Crucially, as none of the transitions between the double states can change both $\indexx$ and $\indexy$ at the same time, the resulting CRN has a bipartite structure.
To fix ideas, the Pac-Man shaped molecule can be thought of as an enzyme binding (state $\indexx$) the monomeric form of a substrate and phosphorylated (state $\indexy$) by adenosine triphosphate ($\F$) to adenosine diphosphate ($\W$) hydrolysis.
A concrete example with a simpler yet similar bipartite structure is found in a recently reported single-molecule molecular motor, where the two states are the axial chirality (which can be in R state or S state) and the chemical state (which can be in diacid state or anhydride state) of the motor~\cite{borsley2022singlebond}.
The latter results in a linear bipartite network to which the treatment in Ref.~\cite{amano2021info} straightforwardly applies.
On the contrary, systems like the present network or the applications analyzed in Section~\ref{sec:applications} require a generalization of the previous framework due to the presence of nonlinear reactions and ancillary species.
}
\label{fig:bipartite}
\end{figure}

We now define the {\color{black}formal} conditions {\color{black} for the chemical species and reactions} under which CRNs have a \textit{bipartite} structure.
{\color{black}An exemplification of the concept is provided in Figure~\ref{fig:bipartite}.}
{\color{black} First, the} chemical species must split as $\setspecies= \setspeciesa \cup \setspeciesb$.
The species $\setspeciesb$ \--- the \textit{bipartite} species \--- are univocally identified with a double \textit{state} $(\indexx, \indexy) = \chemspecies \in \setspeciesb$.
The species $\setspeciesa$ \--- the \textit{ancillary} species \--- are all the other species{\color{black}, when present}.
{\color{black} Second}, the set of chemical reactions must split as $\setrct= \setrcta \cup \setrctx \cup \setrcty$.
The reactions $\setrctx$ (resp. $\setrcty$) interconvert the species $\setspeciesb$ in such a way that only their $\indexx$ (resp. $\indexy$)  state changes:
\begin{subequations}\color{black}
\begin{align}
\dots(\indexx, \indexy)\dots &\ch{->[ ][ $\rct \in \setrctx$ ]} \dots(\indexx', \indexy)\dots\,,\\
\dots(\indexx, \indexy)\dots &\ch{->[ ][ $\rct \in \setrcty$ ]} \dots(\indexx, \indexy')\dots\,.
\end{align}
\end{subequations}
In the above chemical reactions, we neglected the stoichiometric coefficients as well as the other species involved for illustrative reasons. 
The reactions $\setrcta$ are those that do not interconvert the species $\setspeciesb$ ($S^{\chemspecies}_{\rct}=0$ $\forall \chemspecies \in \setspeciesb$ and {\color{black}$\forall \rct\in\setrcta$}):
\begin{equation}\color{black}
\dots(\indexx, \indexy)\dots \ch{->[ ][ $\rct \in \setrcta$ ]} \dots(\indexx, \indexy)\dots\,.
\end{equation}
Notice that{\color{black}, when these conditions for the chemical species and reactions are satisfied,} the species $\setspeciesa$ can undergo every reaction $\rct\in\setrct= \setrcta \cup \setrctx \cup \setrcty$ and no reactions can change both the state $\indexx$ and the state $\indexy$ of a $\setspeciesb$ species.
{\color{black}When these conditions are not satisfied, CRNs do not have a bipartite structure and their splitting into coupled subnetworks (discussed in the next paragraphs of this section) would not be possible.}
{\color{black} Finally, if} the chemical reactions are coarse-grained, we {\color{black}further} assume that the sets $ \setrcta $, $ \setrctx $ and $ \setrcty$ are independent sets of coarse-grained reactions.
The reason for this will become clear in Subs.~\ref{subs:force}.

In bipartite CRNs, we can apply the same splittings of the chemical species and reactions to the stoichiometric matrix
\begin{equation}
\matS = 
 \kbordermatrix{
    & \cr
    \color{g}\setspeciesa &\matSa \cr
    \color{g}\setspeciesb & \matSb \cr
  }\,,
\end{equation}
with 
\begin{subequations}
\begin{align}
\matSa &= 
 \kbordermatrix{
    &\color{g}\setrcta &\color{g}\setrctx &\color{g}\setrcty \cr
    \color{g}\setspeciesa &\matSaa & \matSax & \matSay \cr
  }\,,\\
  \matSb & = 
 \kbordermatrix{
    &\color{g}\setrcta &\color{g}\setrctx &\color{g}\setrcty \cr
    \color{g}\setspeciesb &\matz & \matSbx & \matSby \cr
  }\,,
\end{align}
\end{subequations}
to the current vector
\begin{equation}
\curr = \big(\curra, \currx, \curry \big)\,,
\end{equation}
and to the concentration vector
\begin{equation}
\conc = (\conca,\concb)\,.
\end{equation}
The rate equation~\eqref{eq_dynamics_crns} thus becomes
\begin{subequations}
\begin{align}
&\dt \conca=\matSa  \curr = \matSaa \curra + \matSax \currx + \matSay \curry \label{eq:dynamics_crns_a}\,,\\
&\dt \concb=\matSb  \curr = \matSbx \currx + \matSby \curry \label{eq:dynamics_crns_b}\,.
\end{align}
\end{subequations}

To account for the total concentration of the bipartite species in the same state $\indexx$ or $\indexy$, we define the following marginal concentrations
\begin{subequations}
\begin{align}
[\indexx] &=\sum_{\indexy}[(\indexx,\indexy)]\,,\\
[\indexy] &=\sum_{\indexx}[(\indexx,\indexy)]\,,
\end{align}
\end{subequations}
which are collected in the following vectors
\begin{subequations}
\begin{align}
\concx &=(\dots,[\indexx],\dots)^\intercal\,,\\
\concy &=(\dots,[\indexy],\dots)^\intercal\,.
\end{align}
\end{subequations}
{\color{black}Here, $[(\indexx,\indexy)]$ is the concentration of the specific bipartite species $\chemspecies \in \setspeciesb$ characterized by the double state $(\indexx,\indexy)$, i.e.,  $[(\indexx,\indexy)] = [\chemspecies]_{\chemspecies =(\indexx, \indexy)}$.}

The bipartite structure of a CRN allows us to decompose it into two coupled \textit{subnetworks} $\subx$ and $\suby$.
The subnetwork $\subx$ represents the interconversion of the states $\{\indexx\}$ of the bipartite species driven by the $\setrctx$ reactions.
Analogously, the subnetwork $\suby$ represents the interconversion of the states $\{\indexy\}$ driven by the $\setrcty$ reactions.
Notice that this does not imply that the concentrations $[\indexx]$ (resp. $[\indexy]$) change only because of the reactions $\setrctx$ (resp. $\setrcty$)
as we will see for the example in Subs.~\ref{subs:selfassembly}.


\section{Thermodynamics}
\label{sec:thermo}

Building on the setup of Sec.~\ref{sec:setup}, we introduce the thermodynamic theory of deterministic CRNs~\cite{Rao2016,Avanzini2021,penocchio2021photo} and we specialize it to bipartite CRNs.

\subsection{Chemical Potentials}
\label{sub:chempots}
The free energy contributions carried by the chemical species ${\setspecies}$ in ideal dilute solutions are given by the vector of chemical potentials~\cite{holyst2012}:
\begin{equation}
\chempotential_{\setspecies}= \chempotential^\circ_{\setspecies}+RT\ln\conc\,,
\label{eq:chempotz}
\end{equation}
where $\ln\conc =(\dots,\ln[\chemspecies],\dots)^\intercal$ {\color{black}(hereafter, for every vector $\boldsymbol v = (\dots,v_i,\dots)^\intercal$, $\log \boldsymbol v =(\dots,\log v_i,\dots)^\intercal$)}, $T$ is the temperature of the thermal bath, $R$ is the gas constant, and $\chempotential^\circ_\setspecies$ is the vector of the standard chemical potentials, which in turn are given by the sum of a standard enthalpic $\boldsymbol{h}^\circ_\setspecies$ and entropic $\boldsymbol{s}^\circ_\setspecies$ contributions according to
\begin{equation}
\chempotential^\circ_\setspecies = \boldsymbol{h}^\circ_\setspecies - T \boldsymbol{s}^\circ_\setspecies \, .
\end{equation}

From a thermodynamic standpoint, CRNs are said to be \textit{open} when they exchange free energy with some external reservoirs besides the thermal bath at temperature~$T$.
For instance, \textit{chemicals} can be exchanged with chemostats, i.e., particle reservoirs, and/or \textit{photons} can be exchanged with radiation sources.
Each reservoir is in turn characterized by a chemical potential.

For chemostats, their chemical potentials are given by the vector
\begin{equation}
\chempotential_\setspeciesc = \chempotential^\circ_{\setspeciesc}+RT\ln \concc \,,
\label{eq:chempot_chemostats}
\end{equation}
where $\chempotential^\circ_{\setspeciesc}$ and $\concc$ are the vectors of standard chemical potentials and of the concentrations of the exchanged chemicals, respectively.

For radiation sources, the chemical potential of photons $\gamma_\nu$ at frequency $\nu$ and concentration distribution $n_\nu$ reads~\cite{penocchio2021photo}
\begin{equation}
   \mu_\nu = u_\nu - RT \ln\left(\frac{f_\nu + n_\nu}{n_\nu} \right) \, ,
   \label{eq:mu_nu}
\end{equation}
where $u_\nu = \NA \hbar \omega_\nu$ is the energy carried by a mole of photons at frequency $\nu$ and $f_\nu ={2 \omega_\nu^2}/\NA{\pi c^3_{{}}}$  is the density of photon states at frequency $\nu$ (with $\NA$ the Avogadro's number, $\hbar$ the reduced Planck constant, $c$ the speed of light and $\omega_\nu = 2\pi\nu$ the angular frequency).
When the radiation source is thermal, photons are distributed according to the black body distribution at a certain temperature $T_\mathrm{r}$:
\begin{equation}
   n_\nu^{T_\mathrm{r}} = \frac{f_\nu}{\exp(\hbar\omega_\nu/k_\mathrm{B} T_\mathrm{r}) - 1} \, ,
\end{equation}
with $k_\mathrm{B}$ the Boltzmann constant.
As a consequence, their chemical potential becomes~\cite{Ries1991,Graber1997}
\begin{equation}
   \mu_\nu = u_\nu - RT \ln\left(\frac{f_\nu + n_\nu^{T_\mathrm{r}}}{n_\nu^{T_\mathrm{r}}} \right) = u_\nu \left(1-\frac{T}{T_\mathrm{r}} \right) \, .
   \label{eq:mu_bb}
\end{equation}
Since a radiation source is in equilibrium with a CRN only when the corresponding chemical potential vanishes~\cite{penocchio2021photo}, i.e., $\mu_\nu = 0$, Eq.~\eqref{eq:mu_bb} implies that this happens for a thermal radiation source only when its temperature is the same as the one of the thermal bath, i.e., $T_{\mathrm{r}}=T$.

In this work, we focus for simplicity on autonomous CRNs, namely, the chemical potentials of the reservoirs ($\chempotential_\setspeciesc$ in Eq.~\eqref{eq:chempot_chemostats} for chemostats and $\mu_\nu$ in Eq.~\eqref{eq:mu_nu} for radiation sources), and thus also the quantities $\concc$ and $n_\nu$,  are constant in time.

\subsection{Thermodynamic Forces\label{subs:force}}
Chemical reactions are driven by thermodynamic forces named affinities~\cite{Prigogine1961,Prigogine2015}, 
\begin{equation}
\Aff_\rct = -\frac{\chempotential_{\setspecies}\cdot \colS}{T} + \forces_{\rct} \, ,
\label{eq:aff}
\end{equation}
which have two contributions.
The first contribution, i.e., $-\chempotential_{\setspecies}\cdot \colS/T$, accounts for the variation of the free energy due to the interconversion of the $\setspecies$ species in solution via reaction $\rct$.
The second, i.e., $\forces_{\rct}$, accounts for the \textit{external force} due to the coupling with the reservoirs.

For elementary reactions, affinities~\eqref{eq:aff} satisfy the local detailed balance condition,
\begin{equation}
\Aff_\rct = R\ln\frac{\currs^{+\rct}}{\currs^{-\rct}}\,,
\label{eq:ldt}
\end{equation}
which implies that they {\color{black}always have the same sign of} the corresponding reaction currents: $\Aff_\rct\currs^{\rct}\geq0$.
Coarse-grained reactions do not in general satisfy Eq.~\eqref{eq:ldt}, and hence their affinities may not have the same sign as the currents (unless they are tightly coupled~\cite{Wachtel2018}).
However, every \textit{independent} subset of coarse-grained reactions $\setrctcg\subseteq\setrct$, namely, every subset whose underlying elementary reactions involve unique intermediate (coarse-grained) species which are not shared with other subsets, satisfies $\sum_{\rct\in\setrctcg}\Aff_\rct\currs^{\rct}\geq0$~\cite{Wachtel2018, Avanzini2020b}.

\paragraph*{Example 1.}
Consider the following reaction (assumed as elementary for illustrative reasons) where the catalyst $\mathrm{M}$ interconverts the substrate (fuel) $\F$ into the product (waste) $\W$ which are exchanged with chemostats:
\begin{align}
&\schemestart M\arrow{-U>[$\F$][$\W$]}M\schemestop\,.
\makeatletter
  \chemmove{\node[below] at (Uarrow@arctangent) {}; }
\makeatother
\label{scheme:F2Wcg}
\end{align}
In this case, the affinity coincides with the external force,
\begin{equation}
    \Aff = \forces = \frac{\mu_\F - \mu_\W}{T} \,,
\end{equation}
while the reaction fluxes satisfy mass action kinetics,
\begin{align}
&\currs^{+}=k_{+}[\F][\ch{M}]\,,&\currs^{-}=k_{-}[\W][\ch{M}]\,,
\end{align}
where $[\F]$, $[\W]$ and $[\ch{M}]$ are the concentrations of the fuel, waste and catalyst, respectively.
Since reaction~\eqref{scheme:F2Wcg} is elementary, it satisfies the local detailed balance condition~\eqref{eq:ldt},
which in turn implies that the kinetic constants $\{k_{+}, k_{-}\}$ and the standard chemical potentials $\{\mu_\F^\circ, \mu_\W^\circ\}$ are not independent:
\begin{equation}
\frac{\mu_\F^\circ - \mu_\W^\circ}{T}=R\ln\frac{k_{+}}{k_{-}}\,,
\end{equation}
where we used Eq.~\eqref{eq:chempot_chemostats} to express the chemical potential of the chemostats.

\paragraph*{Example 2.}
Consider the coarse-grained reactions
\begin{align}
\begin{split}
&\schemestart E\arrow{-U>[$\gamma_\nu$]}Z\schemestop\, ,
\makeatletter
\chemmove{\node[below] at (Uarrow@arctangent) {$1$};}
\makeatother
\quad
\schemestart Z\arrow{-U>[$\gamma_{\nu'}$]}E\schemestop\, ,
\makeatletter
\chemmove{\node[below] at (Uarrow@arctangent) {$2$};}
\makeatother
\\
\quad \\
&\schemestart E\arrow{-U>[$\gamma_\nu$]}E\schemestop\, ,
\makeatletter
\chemmove{\node[below] at (Uarrow@arctangent) {$3$};}
\makeatother
\quad
\schemestart Z\arrow{-U>[$\gamma_{\nu'}$]}Z\schemestop\, ,
\makeatletter
\chemmove{\node[below] at (Uarrow@arctangent) {$4$};}
\makeatother
\\
\quad \\
&\qquad \qquad \schemestart E\arrow{-U>[]}Z\schemestop\, ,
\makeatletter
\chemmove{\node[below] at (Uarrow@arctangent) {$5$};}
\makeatother
\end{split}
\label{eq:ex2}
\end{align}
describing a typical photoisomerization process occurring via the so-called diabatic mechanism:
upon the absorption of a photon  $\gamma_\nu$, a species in the $\ch{E}$ conformation is either converted into the $\ch{Z}$ conformation (reaction~$1$) or it decays back to the same conformation by dissipating all the absorbed energy (reaction~$3$);
analogously, another photon $\gamma_{\nu'}$, with possibly a different frequency $\nu'$, can trigger the conversion of $\ch{Z}$ into $\ch{E}$ (reaction~$2$) or be dissipated (reaction~$4$);
finally, $\ch{E}$ and $\ch{Z}$ can also thermally interconvert (reaction~$5$).
We assume for simplicity that only photons at the exact frequencies $\nu$ and $\nu'$ can trigger the corresponding transitions. 
In such a case, the affinities read
\begin{align}\small
    \Aff_{1} = \frac{\mu_\mathrm{E} - \mu_\mathrm{Z}}{T} + \frac{\mu_\nu}{T} \, , \quad \Aff_{2} = \frac{\mu_\mathrm{Z} - \mu_\mathrm{E}}{T} + \frac{\mu_{\nu'}}{T} \, , \nonumber \\ 
    \Aff_{3} = \frac{\mu_\nu}{T} \, , \quad \Aff_{4} = \frac{\mu_{\nu'}}{T} \, , \quad \Aff_{5} = \frac{\mu_\mathrm{E} - \mu_\mathrm{Z}}{T}\,,
\end{align}
with $F_{1} = F_{3} = {\mu_\nu}/{T}$, $F_{2} = F_{4} = {\mu_{\nu'}}/{T}$, and $F_{5} = 0$, while expressions for the reaction currents $\{\currs^1,\currs^2,\currs^3,\currs^4,\currs^5\}$ can be found in Ref.~\cite{penocchio2021photo}.
The reactions~\eqref{eq:ex2} are not tightly coupled and, consequently, affinities and currents do not have, in general, the same sign.
However, as they constitute an independent subset of coarse-grained reactions, they satisfy
\begin{equation}
\Aff_{1} \currs^1+ \Aff_{2} \currs^2+\Aff_{3} \currs^3+ \Aff_{4} \currs^4+\Aff_{5} \currs^5\geq 0\,.
\end{equation}

\subsection{Entropy\label{sub:entropy_def}}
The thermodynamic entropy per unit volume of open CRNs,
\begin{equation}
S = \Sz + \Sch + \Sph\,,
\label{eq:totentropy}
\end{equation}
is given in general by the contribution of the chemical species~$\Sz$, the chemostats~$\Sch$, and the photons~$\Sph$.
The contribution carried by the chemical species reads~\cite{Rao2016}
\begin{equation}
\begin{split}
\Sz
= (\boldsymbol s^\circ_\setspecies-R\ln\conc)\cdot\conc +R\norm{\conc}
= \boldsymbol s^\circ_\setspecies\cdot\conc + \Sh_\setspecies \, ,
\end{split}
\label{eq:entropy_species}
\end{equation}
where $\norm{\conc}=\sum_\chemspecies [\chemspecies]$ 
and we introduced a Shannon-like entropy for macroscopic non-normalized concentration distributions, $\Sh_\setspecies := -R\ln\conc\cdot\conc+R\norm{\conc}$.
The latter will play a crucial role in recovering the information thermodynamic framework for deterministic CRNs in Sec.~\ref{sec:info}.

Similarly, the contribution of the chemostats is given by
\begin{equation}
\Sch= (\boldsymbol s^\circ_\setspeciesc-R\ln\concc)\cdot\concc +R\norm{\concc}\,,
\end{equation}
with $\boldsymbol s^\circ_\setspeciesc= - \partial \chempotential^\circ_{\setspeciesc} /\partial T$ the vector of the standard molar entropies of the chemostats.
On the other hand, the contribution due to the photons is expressed as the entropy of an ideal Bose gas~\cite{Prigogine2015}
\begin{equation}\small
\Sph =  R \int \Big[ \left(f_\nu +  n_\nu\right) \ln \left(f_\nu +  n_\nu\right) - n_\nu \ln n_\nu - f_\nu \ln f_\nu  \Big] \mathrm{d} \nu  \, ,
\label{eq:Sph}
\end{equation}
where the integral runs in general over the whole spectrum.

The time derivative of the total entropy~\eqref{eq:totentropy}, i.e., ${\dt S =\boldsymbol s_\setspecies\cdot\matS\curr}$ (with $\boldsymbol s_\setspecies = \boldsymbol s^\circ_\setspecies-R\ln\conc$), is given by the sum 
of the entropy production rate $\epr\geq0$, accounting for the dissipation,
and of the entropy flow $\efr$, accounting for the reversible exchange of entropy with the reservoirs.
It provides the nonequilibrium formulation of the second law, which can be written as
\begin{align}
\epr = \dt S - \efr \geq0
\label{eq:sh_entropy_dynamics}
\end{align}
where, in autonomous CRNs (i.e., with $\concc$ and $n_\nu$ being constant in time),
\begin{align}
\epr &= \boldsymbol{\Aff}\cdot \curr\geq 0\,,\\
\efr& =\frac{\boldsymbol h^\circ_\setspecies}{T}\cdot\matS\curr - \force\cdot\curr  \,,
\label{eq_efr_def}
\end{align}
with $\boldsymbol{\Aff} = (\dots,\Aff_\rct,\dots)^\intercal$ and $\boldsymbol{\force} = (\dots,\forces_\rct,\dots)^\intercal$.

In bipartite CRNs, we can apply the same splittings of the chemical species $\setspecies = \setspeciesa \cup \setspeciesb$ to the total entropy carried by the chemical species $\Sz$ (given in Eq.~\eqref{eq:entropy_species}) which leads to
\begin{equation}
S = {\Sa + \Sb} + \Sch + \Sph \,,
\end{equation}
where, given a set of species $\setgeneric = \{\setspeciesa,\setspeciesb\}$, $\Sgeneric = (\boldsymbol s^\circ_\setgeneric-R\ln\concgeneric)\cdot\concgeneric +R\norm{\concgeneric}$ with $\concgeneric = (\dots,[\chemspecies],\dots)_{\chemspecies\in\setgeneric}$ and $ \boldsymbol s^\circ_\setgeneric = (\dots, s^\circ_\chemspecies ,\dots)_{\chemspecies\in\setgeneric}$.
Furthermore, we can apply the splitting of the chemical reactions $\setrct= \setrcta \cup \setrctx \cup \setrcty$ to the entropy production rate and the entropy flow:
\begin{align}
\epr &=\epra + \eprx + \epry \label{eq:epr_dec_zero} \,,\\
\efr &= \efra +\efrx + \efry \,,\label{eq:efr_dec}
\end{align}
where
\begin{align}
\epr^\iindex &= \sum_{\rct\in\iindex}{\Aff}_\rct \currs^\rct \geq 0\label{eq:epr_dec_rct}\,,\\
 \efr^\iindex &=  \sum_{\rct\in\iindex} \left(\frac{\boldsymbol h^\circ_\setspecies}{T}\cdot\colS - \forces_\rct \right)\currs^\rct \,,\label{eq:efr_dec_rct}
\end{align}
with $\iindex=\{\setrcta,\setrctx, \setrcty\}$.
Note that the inequality in Eq.~\eqref{eq:epr_dec_rct} holds because the reactions in the sets $\{\setrcta,\setrctx, \setrcty\}$
either satisfy the local detailed balance condition~\eqref{eq:ldt},
or are independent sets of coarse-grained reactions (see Subs.~\ref{sec:setup} and~\ref{subs:etropy_dec}).


\section{Information Thermodynamics}
\label{sec:info}
We now formulate a second law for each subnetwork (defined in Sec.~\ref{sec:setup}) which, compared to the second law~\eqref{eq:sh_entropy_dynamics} of the whole CRN, is (mainly) modified by an information flow term accounting for the information exchange between the two subnetworks.
This information flow arises due to the fact that the two subnetworks share some chemical species which are involved in both $\setrctx$ and $\setrcty$ reactions.

\subsection{Mutual Information}\label{sub:mutual_info}
{\color{black}
We start by introducing the probability that bipartite species $\chemspecies \in \setspeciesb$ are in a specific double state $(\indexx, \indexy)$ as}
\begin{equation}
\pr_{\indexx, \indexy}=\frac{[{\color{black}(}\indexx, \indexy{\color{black})}]}{\nrm}\,,
\label{eq:pxy}
\end{equation}
with $\nrm = \sum_{\chemspecies\in\setspeciesb}[\chemspecies]$ and the marginal probabilities
\begin{subequations}
\begin{align}
\pr_{\indexx} &=\sum_\indexy\pr_{\indexx, \indexy}=\frac{[\indexx]}{\nrm}\,,\\
\pr_{\indexy} &=\sum_\indexx\pr_{\indexx, \indexy}=\frac{[\indexy]}{\nrm}\,,
\end{align}
\label{eq:pxpy}
\end{subequations}
{\color{black}that bipartite species are in state $\indexx$ (resp. $\indexy$) irrespectively of their state $\indexy$ (resp. $\indexx$).}
In analogy to information theory when dealing with the joint {\color{black}probability} of a pair of random variables~\cite{Cover2012}, this allows us to introduce the following Shannon entropies for the bipartite species
\begin{subequations}
\begin{align}
&\Shannon_{\indexx, \indexy} = -\sum_{\indexx, \indexy} \pr_{\indexx, \indexy} \ln\pr_{\indexx, \indexy} \, ,\label{eq:shannon_def_xy} \\
&\Shannon_{\indexx} = -\sum_{\indexx} \pr_{\indexx} \ln\pr_{\indexx} \, , \\
&\Shannon_{\indexy} = -\sum_{\indexy} \pr_{\indexy} \ln\pr_{\indexy} \, ,
\end{align}
\label{eq:shannon_def}
\end{subequations}
and the \textit{mutual information} 
\begin{equation}
\begin{split}
\info
= \sum_{\indexx, \indexy}\pr_{\indexx, \indexy} \ln\frac{\pr_{\indexx, \indexy}}{\pr_{\indexx}\pr_{\indexy}}\,,
\end{split}
\label{eq:info_def_standard_infotheory}
\end{equation}
which by construction satisfies
\begin{equation}
    \Shannon_{\indexx, \indexy} = \Shannon_{\indexx} + \Shannon_{\color{black}\indexy} - \info \, .
    \label{eq:standard_Shannon_decomposition}
\end{equation}
Crucially, by applying the definition of probabilities in Eqs.~\eqref{eq:pxy} and~\eqref{eq:pxpy}, the mutual information~\eqref{eq:info_def_standard_infotheory} can be directly expressed in terms of the macroscopic non-normalized concentration distributions
\begin{equation}
\info =\sum_{\indexx, \indexy}\frac{[{\color{black}(}\indexx, \indexy{\color{black})}]}{\nrm} \ln\frac{[{\color{black}(}\indexx, \indexy{\color{black})}]\nrm}{[\indexx][\indexy]} \, ,
\label{eq:info_def}
\end{equation}
{\color{black}which still measures, as in information theory~\cite{parrondo2015,Cover2012}, the reduction in uncertainty about the states $\{\indexx\}$ when the states $\{\indexy\}$ are known, and vice versa.
This physically means that $\info$ quantifies, in the framework of this paper, to which extent being in a specific state $\indexx$ correlates with being in a specific state $\indexy$ for the bipartite species and it can be determined by measuring the concentrations $[(\indexx,\indexy)]$.}

{\color{black}\paragraph*{Example.} 
Consider the network in Fig.~\ref{fig:bipartite} with arbitrary concentrations $[(\indexx,\indexy)]$.
On the one hand, if $[(\indexx_1,\indexy_1)]/[(\indexx_2,\indexy_1)] = [(\indexx_1,\indexy_2)]/[(\indexx_2,\indexy_2)]$, the $\indexy$ state is not correlated with the $\indexx$ state, as the relative concentration between species in states $\indexx_1$ and $\indexx_2$ is independent of state $\indexy$.
Analogously, if $[(\indexx_1,\indexy_1)]/[(\indexx_1,\indexy_2)] = [(\indexx_2,\indexy_1)]/[(\indexx_2,\indexy_2)]$, the $\indexx$ state is not correlated with the $\indexy$ state, as the relative concentration between species in states $\indexy_1$ and $\indexy_2$ is independent of state $\indexx$.
In such a case, one cannot gain information about one state (e.g., the concentration of molecules binding the triangular substrate) by measuring the other (e.g., the concentration of molecules binding the small circular substrate).
This reflects into a vanishing mutual information~\eqref{eq:info_def}.
On the other hand, if $[(\indexx_1,\indexy_1)]/[(\indexx_2,\indexy_1)] \neq [(\indexx_1,\indexy_2)]/[(\indexx_2,\indexy_2)]$, the $\indexy$ state correlates with the $\indexx$ state, as the relative concentration between species in states $\indexx_1$ and $\indexx_2$ now depends state $\indexy$.
For analogous reasons,  if $[(\indexx_1,\indexy_1)]/[(\indexx_1,\indexy_2)] \neq [(\indexx_2,\indexy_1)]/[(\indexx_2,\indexy_2)]$, the $\indexx$ state correlates with the $\indexy$ state.
In such a case, one can thus gain information on one state by measuring the other.
For instance, in the presence of positive correlations between $\indexx_1$ and $\indexy_1$ and between $\indexx_2$ and $\indexy_2$ (namely, $[(\indexx_1,\indexy_1)] > [(\indexx_1,\indexy_2)]$ and $[(\indexx_2,\indexy_2)] > [(\indexx_2,\indexy_1)]$), knowing that the marginal concentration $[\indexy]$ is shifted towards state $\indexy_2$ (most of the molecules bind the small circular substrate) informs about the marginal concentration $[\indexx]$ being shifted towards state $\indexx_2$ (most of the molecules bind the triangular substrate).
This reflects into a non-null mutual information~\eqref{eq:info_def}.}


\subsection{Entropy Decomposition\label{subs:etropy_dec}}
{\color{black}
In standard information thermodynamics, the thermodynamic entropy of a system coincides with the Shannon entropy.}
Thus, Eq.~\eqref{eq:standard_Shannon_decomposition} splits the {\color{black} former} into contributions from subsystems and their mutual information.
{\color{black} On the other hand, as} shown in Subs.~\ref{sub:entropy_def}, the entropy~\eqref{eq:entropy_species} of deterministic CRNs is not given by the Shannon entropy.
{\color{black} Nevertheless}, the bipartite structure of the network allows us to split the Shannon-like entropy of the bipartite species $\Shb = -R(\ln\concb)\cdot\concb+R\nrm$ {\color{black} by following the same logic employed for the standard derivation of Eq.~\eqref{eq:standard_Shannon_decomposition}~\cite{Cover2012}, but exploiting the expression in Eq.~\eqref{eq:info_def} for the mutual information.
In this way, we get}
\begin{equation}
\Shb=\Shx+\Shy - R\nrm \info + R\nrm(\ln\nrm - 1)\,,
\label{eq:decomp_shannon_entropy}
\end{equation}
where
\begin{subequations}
\begin{align}
\Shx & = -R(\ln\concx)\cdot\concx+R\norm{\concx}\,,\label{eq:shannon_x}\\
\Shy & = -R(\ln\concy)\cdot\concy+R\norm{\concy}\,,\label{eq:shannon_y}
\end{align}
\end{subequations}
are the Shannon-like entropy of states $\{\indexx\}$ and $\{\indexy\}$, respectively, and $\norm{\concx}=\norm{\concy}=\norm{\concb}$.
Note that the last term in Eq.~\eqref{eq:decomp_shannon_entropy}, i.e., $\nrm(\ln\nrm - 1)$, emerges because of the non-normalized concentration distributions.

By using Eq.~\eqref{eq:decomp_shannon_entropy} to express the entropy of the bipartite species $\Sb = \Shb + \boldsymbol s^\circ_\setspeciesb\cdot\concb$, the total entropy~\eqref{eq:totentropy} becomes 
\begin{equation}
\begin{split}
S = \text{ }
&\Shx + \Shy - R\nrm \info + R\nrm(\ln\nrm - 1) + \boldsymbol s^\circ_{\concb}\cdot\concb \\
& +\Sa + \Sch + \Sph \,,
\end{split}
\label{eq_shannon_entropy_info}
\end{equation}
where $ \boldsymbol s^\circ_{\concb}\cdot\concb$ accounts for the standard molar entropy of the bipartite species and cannot be split into a contribution due to the species in state $\indexx$ and a contribution due to the species in state $\indexy$.
This term is absent in the decomposition of the entropy of bipartite Markov jump processes~\cite{horowitz2014} as long as the states do not have an internal entropy.

\subsection{Entropy Production Decomposition}
We now decompose the entropy production rate in such a way as to formulate the second law for each subnetwork $\subx$ and $\suby$.
We start by rewriting the second law~\eqref{eq:sh_entropy_dynamics} specifying the entropy flow according to Eq.~\eqref{eq:efr_dec}:
\begin{equation}
\epr = \dt S - \efra - \efrx - \efry \,.
\label{eq_entropy_decomp_1}
\end{equation}
From Eq.~\eqref{eq_shannon_entropy_info}, we further derive 
\begin{equation}
\begin{split}
\dt S =& 
\dt\Shx+\dt\Shy + \boldsymbol s^\circ_{\concb}\cdot\dt\concb - R\dt\big(\nrmt \info\big) + \\
&+ R\ln\nrmt  \dt\nrmt + \boldsymbol s_{\setspeciesa} \cdot\dt\conca\, ,
\end{split}
\label{eq_entropy_dynamics_1}
\end{equation}
where $\boldsymbol s_{\setspeciesa} = \boldsymbol s_{\setspeciesa}^\circ - R\ln\conca$ and the time derivative of the information term~\eqref{eq:info_def} can be specified as 
\begin{equation}
\dt \info = \infoflowx + \infoflowy + \big(\ln\nrmt-\info\big){\dt\ln\nrmt}
\label{eq:info_dynamics_1}
\end{equation}
with
\begin{subequations}
\begin{align}
& \infoflowx :=\frac{1}{\nrmt}\Big\{ (\ln\concb)\cdot\matSbx \currx - (\ln\concx)\cdot\dt\concx\Big\}\,,\label{eq:inf_flow_x}\\
& \infoflowy :=\frac{1}{\nrmt}\Big\{ (\ln\concb)\cdot\matSby \curry - (\ln\concy)\cdot\dt\concy\Big\}\,.\label{eq:inf_flow_y}
\end{align}
\end{subequations}
Crucially, the information flows $\infoflowx$ and $\infoflowy$ quantify the changes in the mutual information~\eqref{eq:info_def} per units of concentration,
namely, the changes in the relative uncertainty (correlations) between the $\{\indexx\}$ and the $\{\indexy\}$ states, due to the dynamics of subnetworks $\subx$ and $\suby$, respectively.
The third term in Eq.~\eqref{eq:info_dynamics_1} accounts for the non-normalized nature of the concentration distributions, vanishes at steady state, and does not contribute to the entropy production decomposition.
Indeed, by using Eq.~\eqref{eq_entropy_dynamics_1} and Eq.~\eqref{eq:info_dynamics_1} into Eq.~\eqref{eq_entropy_decomp_1} and by splitting $\dt\conca$ and $\dt\concb$ into the contributions due to the different reactions $\setrct= \setrcta \cup \setrctx \cup \setrcty$ as in Eqs.~\eqref{eq:dynamics_crns_a} and~\eqref{eq:dynamics_crns_b}, we obtain 
{
\begin{equation}
\begin{split}
\epr(t)=
&{\dt\Shx - \efrx - R\nrmt \infoflowx + \boldsymbol s^\circ_{\concb}\cdot \matSbx \currx + \boldsymbol s_{\setspeciesa}\cdot\matSax \currx}+\\
& +{\dt\Shy - \efry - R\nrmt \infoflowy + \boldsymbol s^\circ_{\concb}\cdot \matSby \curry + \boldsymbol s_{\setspeciesa}\cdot\matSay \curry}+\\
& -{\efra + \boldsymbol s_{\setspeciesa}\cdot\matSaa \curra}
\end{split}
\label{eq:epr_slitted}
\end{equation}}
which, by using Eq.~\eqref{eq:epr_dec_rct}, allows us to identify 
{\small
\begin{subequations}
\begin{align}
\eprx & = \dt\Shx - \efrx - R\nrmt \infoflowx + \boldsymbol s^\circ_{\concb}\cdot \matSbx \currx + \boldsymbol s_{\setspeciesa}\cdot\matSax \currx \geq0\label{eq:final_deco_x}\,, \\
\epry & = \dt\Shy - \efry - R\nrmt \infoflowy + \boldsymbol s^\circ_{\concb}\cdot \matSby \curry + \boldsymbol s_{\setspeciesa}\cdot\matSay \curry \geq0 \label{eq:final_deco_y}\,, \\
\epra & = -\efra + \boldsymbol s_{\setspeciesa}\cdot\matSaa \curra\geq0 \,. \label{eq:final_deco_a}
\end{align}\label{eq:final_deco}
\end{subequations}}
Equation~\eqref{eq:final_deco_a} has the same form as the second law~\eqref{eq:sh_entropy_dynamics}:
it explicitly expresses the dissipation of the $\setrcta$ reactions (which do not involve the bipartite species) in terms of the entropy flow with the corresponding reservoirs $\efra$ and the variation of the total entropy due to the interconversion of the $\setspeciesa$ species $\boldsymbol s_{\setspeciesa}\cdot\matSaa \curra$.

Equations~\eqref{eq:final_deco_x} and~\eqref{eq:final_deco_y} are the key results of our work. 
They provide a formulation of the second law for the subnetworks $\subx$ and $\suby$, respectively.
Let us focus in the following on the subnetwork $\subx$, but analogous comments also hold for the subnetwork $\suby$. 
Equation~\eqref{eq:final_deco_x} shows that the dissipation $\eprx$ is balanced by different mechanisms:
i) the variation of the subnetwork Shannon-like entropy,~$\dt\Shx$;
ii) the entropy flow with the reservoirs coupled to the $\setrctx$ reactions,~$\efrx$;
iii) the information flow accounting for the information exchange with the $\suby$ subnetwork,~$\infoflowx$;
iv) the variation to the subnetwork entropy due to the standard molar entropy carried by the chemical species,~$\boldsymbol s^\circ_{\concb}\cdot \matSbx \currx$;
v) the variation to the subnetwork entropy due to the consumption/production of the $\setspeciesa$ species,~$\boldsymbol s_{\setspeciesa}\cdot\matSax \currx$.
Note that the first three mechanisms in Eq.~\eqref{eq:final_deco_x} also appear in an analogous decomposition of entropy production for bipartite Markov jump processes~\cite{horowitz2014}, while the last two only emerge in bipartite deterministic CRNs.

Each mechanism can act as a \textit{source} (of entropy), when it contributes positively to the entropy production, or as an \textit{output} (of entropy), when it contributes negatively.
The sources must always dominate the outputs to maintain the subnetwork out of equilibrium because of the unavoidable dissipation.
For instance, when $\infoflowx < 0$, the subnetwork $\subx$ is decreasing correlation with the subnetwork $\suby$:
the mutual information~\eqref{eq:info_def} decreases (see Eq.~\eqref{eq:info_dynamics_1}).
This consumption of mutual information corresponds to a source of entropy ($- R\nrmt \infoflowx > 0$) that can be used to sustain the output mechanisms and/or to balance the dissipation.
On the other hand, when $\infoflowx > 0$ the subnetwork is increasing correlation with the subnetwork $\suby$.
This corresponds to an output of entropy ($- R\nrmt \infoflowx < 0$) which has to be sustained by an entropy source to balance the dissipation ($\eprx \geq 0$).


\section{Applications}
\label{sec:applications}

We now apply our framework to two paradigmatic examples of chemical engines with nonlinear dynamics: a model for chemically-driven self-assembly~\cite{ragazzon2018,prins2021} and a light-driven bimolecular motor~\cite{ragazzon2015,sabatino2019,credi2021,asnicar2022}.
We characterize their functioning at steady state for various operating regimes.

\subsection{Driven self-assembly}
\label{subs:selfassembly}

\begin{figure*}[t]
        \includegraphics[width=.98\textwidth]{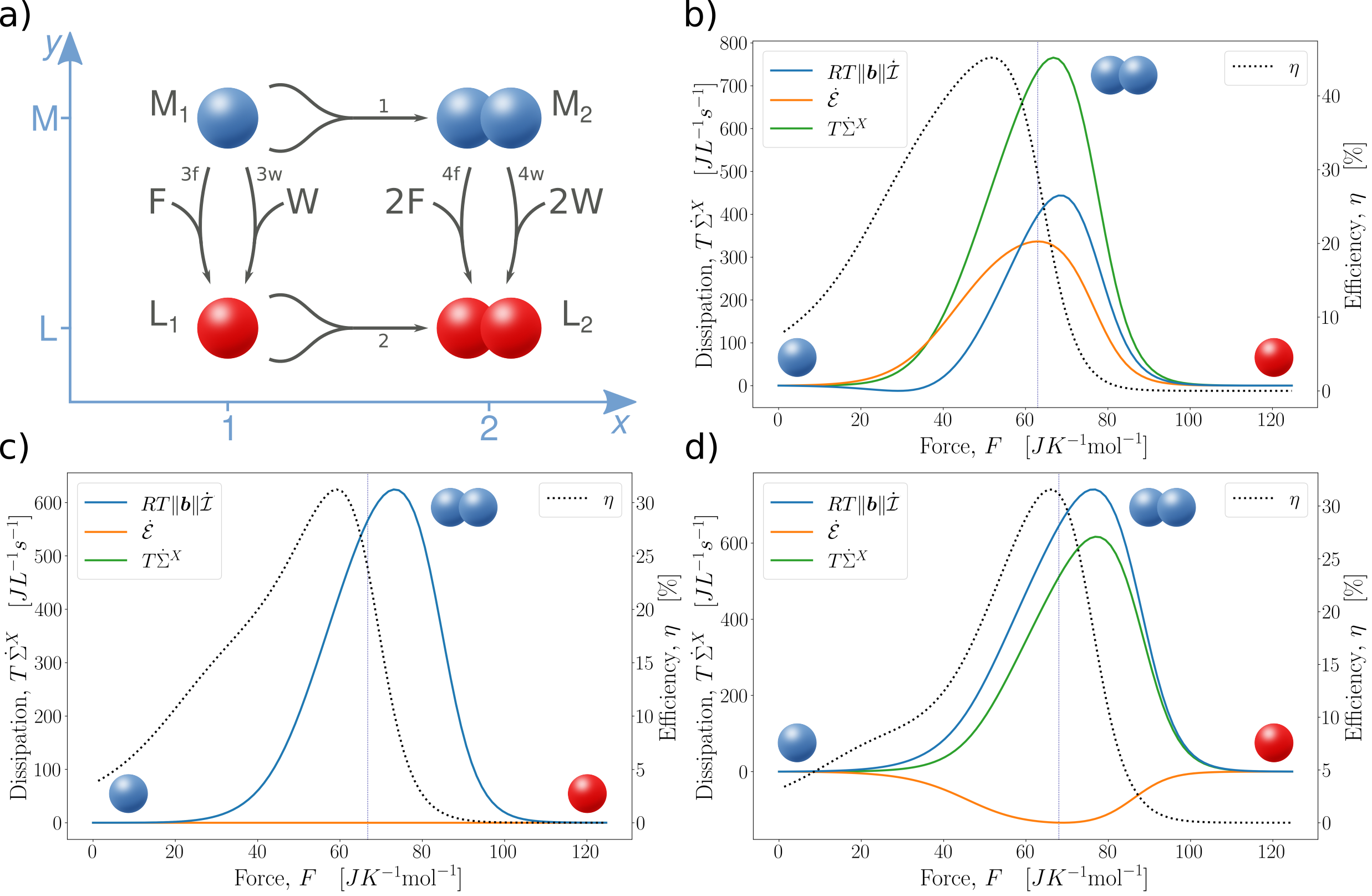}
        \centering
        \caption{
            \textbf{Chemically-driven self-assembly of monomers into dimers.}
a) Hypergraph representation of the bipartite chemical reaction network.
Arrows are used to indicate the conventional direction of the reactions, which are all reversible.
Reactions are labelled according to the CRN~\eqref{rct:drivenSA};
b-d)  Numerical simulations of the model with three different sets of parameters. 
Information flow ($\IF$), energy flow ($\EF$), and dissipation of the self-assembly subnetwork ($T\eprx = \EF + \IF$) at steady state are plotted against the net thermodynamic force acting on the system ($F = ({\mu_\mathrm{F} - \mu_\mathrm{W}}) / T$, with $T = 298$K, $\mu^\circ_\F = 11$ kJ{\color{black}/mol} and $\mu^\circ_{\color{black}\W} = -11$ kJ{\color{black}/mol, up to an arbitrary constant}) in a range corresponding to $[\mathrm{F}]$ from $1\cdot10^{-4}$M to $4\cdot10^{2}$M, with $[\mathrm{W}] = 1$M.
The efficiency of the internal free-energy transduction ($\eta$) is computed according to Eq.~\eqref{eq:effsa}.
The standard chemical potentials ruling the relative thermodynamic stability of the four species are{\color{black}, up to an arbitrary constant}: $\mu^\circ_{\mon_1} = -2$ kJ{\color{black}/mol}, $\mu^\circ_{\mona_1} = -3$ kJ{\color{black}/mol}, $\mu^\circ_{\mona_2} = -4$ kJ{\color{black}/mol}, and $\mu^\circ_{\mon_2} = 9$ kJ{\color{black}/mol} for case b); $\mu^\circ_{\mon_1} = 2$ kJ{\color{black}/mol}, $\mu^\circ_{\mona_1} = -3$ kJ{\color{black}/mol}, $\mu^\circ_{\mona_2} = -6$ kJ{\color{black}/mol}, and $\mu^\circ_{\mon_2} = 4$ kJ{\color{black}/mol}  for case c); and  $\mu^\circ_{\mon_1} = 2$ kJ{\color{black}/mol}, $\mu^\circ_{\mona_1} = -3$ kJ{\color{black}/mol}, $\mu^\circ_{\mona_2} = -2$ kJ{\color{black}/mol}, and $\mu^\circ_{\mon_2} = 4$ kJ{\color{black}/mol} for case d).
The independent kinetic parameters are the same as in Ref.~\cite{penocchio2019eff}.
Colored spheres indicate which is the most populated species at steady state for low, intermediate and high forces regimes, while dotted vertical lines mark the value of the force maximizing the concentration of the target species $\mon_2$.
{\color{black}Note that, in panel c), the information flow ($\IF$, blue solid line) coincides exactly with the dissipation of the self-assembly subnetwork ($T\eprx$, green solid line).}
        }
        \label{fig:drivenSA}
\end{figure*}

As a first application, we analyze a minimalist model (see Fig.~\ref{fig:drivenSA}a) epitomizing the basic working principles of chemically-driven self-assembly~\cite{ragazzon2018,falasco2019ndr,penocchio2019eff,prins2021}:
an external force is exploited to increase the concentration of a target species with respect to the equilibrium one.
Prominent examples of this kind of mechanism are the formation of microtubules out of tubulin dimers fueled by guanosine 5'-triphosphate (GTP)~\cite{desai1997,ross2017} and the ATP-driven self-assembly of actin ﬁlaments~\cite{Howard2001}.
Driven self-assembly has also been exploited in experiments such as the controlled gelation of dibenzoyl-L-cysteine to form nanoﬁbers~\cite{boekhoven2010} and the chemically fueled transient self-assembly of ﬁbrous hydrogel materials~\cite{boekhoven2015}.

In the model depicted in Fig.~\ref{fig:drivenSA}a, the direct aggregation of two monomers $\mon_1$ to form the dimer $\mon_2$ is coupled with the exergonic conversion of a high energetic species \ch{F} into a low energetic one \ch{W}.
In particular, both the monomer $\mon_1$ and the dimer $\mon_2$ can catalyze the F-to-W conversion via their activated species $\mona_1$ and $\mona_2$ (e.g., $\F + \mon_1 \ch{->} \mona_1 \ch{->} \mon_1 + \W$).
This leads, by properly fixing the concentrations of \ch{F} and \ch{W}, to a nonequilibrium steady state enriched in the dimer $\mon_2$.
Unlike conventional equilibrium self-assembly, the efficacy of this synthetic procedure is not determined by the relative thermodynamic stability of the species,
but rather by a kinetic asymmetry~\cite{astumian1996} in how the monomers $\mon_1$ and the dimer $\mon_2$ react with $\F$ and $\W$~\cite{ragazzon2018,prins2021}.
This built-in kinetic asymmetry is often referred to as an information ratchet mechanism {\color{black} because the rate at which $\F$ and $\W$ react with the system is somehow dependent on information about the reacting species: that is, a species will react more quickly with fuel if doing so enables forward cycling or prevents backward cycling}~\cite{ragazzon2018,astumian2019}.


The model depicted in Fig.~\ref{fig:drivenSA}a corresponds to the open CRN
\begin{equation}
\begin{split}
2 \mathrm{\mon_1} \ch{->[ $\mathrm{ }$ ][ $1$ ]} \mathrm{\mon_2} \quad\quad\quad\quad
2 \mathrm{\mona_1} \ch{->[ $\mathrm{ }$ ][ $2$ ]} \mathrm{\mona_2}&   \\
\quad \\
\schemestart $\mon_1$\arrow{-U>[$\F$][]}$\mona_1$\schemestop
\makeatletter
\chemmove{\node[below] at (Uarrow@arctangent) {\footnotesize{$3{\color{black}\mathrm{f}}$}}; }
\makeatother \, \quad 
\schemestart $\mon_1$\arrow{-U>[$\W$][]}$\mona_1$\schemestop
\makeatletter
\chemmove{\node[below] at (Uarrow@arctangent) {\footnotesize{$3{\color{black}\mathrm{w}}$}}; }
\makeatother& \\
\quad \\
\schemestart $\mon_2$\arrow{-U>[$2\F$][]}$\mona_2$\schemestop
\makeatletter
\chemmove{\node[below] at (Uarrow@arctangent) {\footnotesize{$4{\color{black}\mathrm{f}}$}}; }
\makeatother \, \quad 
\schemestart $\mon_2$\arrow{-U>[$2\W$][]}$\mona_2$\schemestop
\makeatletter
\chemmove{\node[below] at (Uarrow@arctangent) {\footnotesize{$4{\color{black}\mathrm{w}}$}}; }
\makeatother&
\end{split}
\label{rct:drivenSA}
\end{equation}
where there are four bipartite species  $\setspeciesb = \{\mon_1,\mon_2,\mona_1,\mona_2 \}$ (with monomeric/dimeric states $\indexx \in \{1,2\}$ and non-activated/activated states $\indexy\in\{\mon,\mona\}$) and two chemostats $\F$ and $\W$.
No $\setspeciesa$ species are present.
{\color{black}
The marginal concentrations read
\begin{subequations}
\begin{align}
[\mon] &=[\mon_1] + [\mon_2]\,,\\
[\mona] &=[\mona_1] + [\mona_2]\,,\\
[1] &=[\mon_1] + [\mona_1]\,,\\
[2] &=[\mon_2] + [\mona_2]\,,
\end{align}
\label{eq:marginalDS}
\end{subequations}}
Here, all the reactions satisfy mass action kinetics.
The self-assembly reactions $\{ 1, 2 \}$ are nonlinear with respect to the chemical species $\mon_1$ and $\mona_1$, and only change their $\indexx$ state, i.e., $\setrctx = \{ 1, 2\}$.
They thus define the self-assembly subnetwork $\subx$.
Conversely, the fueling and waste-forming reactions $ \{ 3{\color{black}\mathrm{f}}, 3{\color{black}\mathrm{w}}, 4{\color{black}\mathrm{f}}, 4{\color{black}\mathrm{w}} \}$, accounting for the coupling with the chemostats, only change the $\indexy$ state of the species, i.e., $\setrcty = \{ 3{\color{black}\mathrm{f}}, 3{\color{black}\mathrm{w}}, 4{\color{black}\mathrm{f}}, 4{\color{black}\mathrm{w}} \}$.
They thus define the fueling subnetwork $\suby$.
{\color{black}We stress that here, due to the nonlinearity of the $\setrctx$ reactions, $\nrm = [\mon] + [\mona] = [1] + [2] = [\mon_1] + [\mon_2] + [\mona_1] + [\mona_2]$ is not a conserved quantity of the dynamics, which prevents direct application of previous formulations of information thermodynamics for bipartite networks~\cite{horowitz2014,amano2021info}.
In addition, the CRN in Eq.~\ref{rct:drivenSA} provides an example where the marginal concentrations of $\indexy$ states ($[\mon]$ and $[\mona]$) change due to $\setrctx$ reactions ($\{ 1, 2 \}$).}

We now characterize the energetics of the nonequilibrium steady state which will eventually be reached by the dynamics in the long time limit.
We do so by computing the various contributions to the total dissipation~\eqref{eq:epr_slitted}{\color{black}, thus specializing Eqs.~\eqref{eq:final_deco_x} and~\eqref{eq:final_deco_y} for the case at study (notice that Eq.~\eqref{eq:final_deco_a} as well as all the terms involving $\setspeciesa$ species do not play any role here)}.
At steady state, {\color{black}the time derivatives of Shannon-like entropies of states $\{\indexx\}$ and $\{\indexy\}$ vanish ($\dt\Shx=\dt\Shy=0$)} and the dynamics is fully characterized by the net current $\currs = - \currs^1 = \currs^2 = (\currs^{3{\color{black}\mathrm{f}}} + \currs^{3{\color{black}\mathrm{w}}}) / 2 = -(\currs^{4{\color{black}\mathrm{f}}} + \currs^{4{\color{black}\mathrm{w}}})$ {\color{black}, which is} positive when flowing anticlockwise across the hyper-graph in Fig.~\ref{fig:drivenSA}a {\color{black}according to the sign convention in Eq.~\eqref{rct:drivenSA}.
Furthermore, the rate at which the fuel species $\F$ is injected into the system at steady state to keep its concentration constant ($\excurrs_\F = \currs^{3{\color{black}\mathrm{f}}} + 2\currs^{4{\color{black}\mathrm{f}}}$) corresponds to the rate at which the waste species $\W$ is extracted ($\excurrs_\F = -\excurrs_\W = -\currs^{3\mathrm{w}} - 2\currs^{4\mathrm{w}}$)}.
As a consequence, Eqs.~\eqref{eq:final_deco_x} and~\eqref{eq:final_deco_y} boil down to:
{\small
\begin{subequations}
\begin{align}
T\eprx & = \underbrace{\currs(\mu^\circ_\mathrm{\mon_2} - \mu^\circ_\mathrm{\mona_2} + 2\mu^\circ_\mathrm{\mona_1} - 2\mu^\circ_\mathrm{\mon_1})}_{=:\dot{\mathcal{E}}} + \underbrace{RT\currs \ln \frac{[\mathrm{\mona_1}]^2[\mathrm{\mon_2}]}{[\mathrm{\mon_1}]^2 [\mathrm{\mona_2}]}}_{=:RT\nrmt\dot{\mathcal{I}}} \label{eq:eprDSx}\,, \\
T\epry & = \excurrs_\F (\mu_\F - \mu_\W) - \dot{\mathcal{E}} - RT\nrmt\dot{\mathcal{I}} \label{eq:eprDSy}\,,
\end{align}\label{eq:eprDS}
\end{subequations}
}
where $\dot{\mathcal{I}} =- \infoflowx = \infoflowy$ {\color{black}follows directly from Eqs.~\eqref{eq:inf_flow_x}, \eqref{eq:inf_flow_y} and~\eqref{eq:marginalDS}}, and we introduced the energy flow $\dot{\mathcal{E}} / T=-\efrx + \boldsymbol s^\circ_{\concb}\cdot \matSbx \currx$ accounting for the standard contribution to the variation of the CRN free energy due to $\setrctx$ reactions{\color{black}, with $\excurrs_\F (\mu_\F - \mu_\W) - \dot{\mathcal{E}} = -\efry + \boldsymbol s^\circ_{\concb}\cdot \matSby \curry$ following directly from Eq.~\eqref{eq:efr_dec_rct} (after identifying $\forces_\mathrm{3f} = \mu_\F/T$, $\forces_\mathrm{3w} = \mu_\W/T$, $\forces_\mathrm{4f} = 2\mu_\F/T$, and $\forces_\mathrm{4w} = 2\mu_\W/T$)}.

Equations~\eqref{eq:eprDSx} and~\eqref{eq:eprDSy} fully characterize the free-energy exchanges between the CRN and the chemostats, on the one hand, and between the subnetworks $\subx$ and $\suby$, on the other hand.
The former is accounted for by $\excurrs_\F (\mu_\F - \mu_\W)$, quantifying the fueling work performed by the external force $F = (\mu_\F - \mu_\W) / T$~\cite{penocchio2019eff}.
This work is entirely delivered to the subnetwork $\suby$ since $\excurrs_\F (\mu_\F - \mu_\W)$ appears only in Eq.~\eqref{eq:eprDSy}.
The free-energy exchange between  $\suby$ and $\subx$ is accounted for by the energy and information flows, which together ($\EF + \IF$) constitute the only possible source of free energy for the subnetwork $\subx$ (see Eq.~\eqref{eq:eprDSx}).
This implies that the subnetwork $\subx$ can be kept out of equilibrium only when part of the work $\excurrs_\F (\mu_\F - \mu_\W)$ is not dissipated by the subnetwork $\suby$, but output to the subnetwork $\subx$ via $\dot{\mathcal{E}}$ and $\dot{\mathcal{I}}$.
We view this process as an internal free-energy transduction.
Based on this understanding, one can define the efficiency of the internal free-energy transduction as the ratio between the free energy transferred to the subnetwork~$\subx$ and the work performed on the subnetwork~$\suby$
\begin{align}
   0 \leq \eta = \frac{RT\nrmt\dot{\mathcal{I}} + \dot{\mathcal{E}}}{{I_F}(\mu_\F - \mu_\W)} = 1-\frac{T\epry}{{I_F}(\mu_\F - \mu_\W)} \leq 1 \,,
   \label{eq:effsa}
\end{align}
which is bounded by zero and one due to the non-negativity of~$\epry$.
Physically, $\eta$ measures the fraction of the fueling work devoted to keep the subnetwork $\subx$ out of equilibrium, which is exactly the goal of driven self-assembly.

Such a level of resolution on how free energy is dissipated leads to a refinement of a previous analysis of this model~\cite{penocchio2019eff}.
First, Eq.~\eqref{eq:effsa} characterizes the steady state performance of the self-assembly (what is called ``maintenance phase'' in Ref.~\cite{penocchio2019eff}).
Second, if the subnetwork $\subx$ performed work $\mathcal{W}_{ext}<0$ against the environment (as in the driven synthesis setup of Ref.~\cite{penocchio2019eff}), Eq.~\eqref{eq:effsa} would allow us to split the overall efficiency $\eta_\mathrm{ds} = -\dot{\mathcal{W}}_{ext}/\mathrm{I_F}(\mu_\F - \mu_\W)$ into two contributions:
\begin{align}
  \eta_\mathrm{ds} = \frac{RT\nrmt\dot{\mathcal{I}} + \dot{\mathcal{E}}}{{I_F}(\mu_\F - \mu_\W)} \cdot \frac{-\dot{\mathcal{W}}_{ext}}{\dot{\mathcal{E}} + RT\nrmt\dot{\mathcal{I}}} =\eta \cdot \eta_{ext} \leq 1 \, ,
\end{align}
where $\eta_{ext}$ measures the fraction of the free-energy transduced towards the subnetwork $\subx$ ($\dot{\mathcal{E}} + RT\nrmt\dot{\mathcal{I}}$) which is fruitfully converted into useful power delivered to the environment ($-\dot{\mathcal{W}}_{ext}$).

As a further comment, we note that $\eta$ can vanish because of two reasons: a thermodynamic and a kinetic one.
The former occurs when the external force is null ($F =  (\mu_\F - \mu_\W) /T = 0$), which directly implies that both subnetworks reach equilibrium ($ \epry = \eprx = 0$).
The latter occurs when the external force is not null ($F = (\mu_\F - \mu_\W) /T \neq 0$), but the sum of  $\dot{\mathcal{E}}$ and $RT\nrmt\dot{\mathcal{I}}$ vanishes, thus implying $\eprx = 0$, but $ \epry > 0$.
From a kinetic standpoint, this second case is often referred as a scenario where the system does not display kinetic asymmetry~\cite{ragazzon2018,astumian2019,prins2021}.
From an information thermodynamic standpoint, this happens when there is no internal free-energy transduction between the two subnetworks and, consequently, the subnetwork $\subx$ remains at equilibrium even though $\suby$ is maintained in a nonequilibrium steady state by the chemostats.
This also implies that all the dissipation happens at the level of the subnetwork $\suby$ ($\epr = \epry$).

To better illustrate our results, we examine three different scenarios.
The first one (Fig.~\ref{fig:drivenSA}b) is the same as in Refs.~\cite{penocchio2019eff} and~\cite{falasco2019ndr},
where the dimeric state is energetically highly unfavored in the non-activated state ($\mu^\circ_{\mon_2} - 2\mu^\circ_{\mon_1} \gg 0$), but favored in the activated state ($\mu^\circ_{\mona_2} - 2\mu^\circ_{\mona_1} < 0$).
Under these conditions, the energy flow is positive for any value of the external force and has a maximum at $F=63$ J K$^{-1}$mol$^{-1}$ (which also corresponds to the value of the force maximizing the concentration of $\mon_2$).
This happens because $\EF$ given in Eq.~\eqref{eq:eprDS} is proportional to i) the steady state current $\currs>0$ and ii) $(\mu^\circ_\mathrm{\mon_2} - \mu^\circ_\mathrm{\mona_2} + 2\mu^\circ_\mathrm{\mona_1} - 2\mu^\circ_\mathrm{\mon_1})>0$.
Thus, the energy flow always contributes positively to the internal free-energy transduction from the subnetwork $\suby$ to the subnetwork $\subx$.
In contrast, the information flow $\dot{\mathcal{I}}$ is negative for small values of the external force ($F\lesssim$ 35 J K$^{-1}$mol$^{-1}$), thus representing a cost for the free-energy transduction.
Physically, this means that the subnetwork $\subx$ is spending a fraction of the free energy provided by the energy flow to create correlations with the subnetwork $\suby$.
For intermediate values of the external force (35 J K$^{-1}$mol$^{-1}$ $\lesssim F \lesssim$ 100 J K$^{-1}$mol$^{-1}$), the information flow has a finite positive value meaning that correlations are generated by the subnetwork $\suby$ and used as a source by the subnetwork $\subx$ to stay out of equilibrium. 

In the second scenario (Fig.~\ref{fig:drivenSA}c), there is no energetic preference between monomeric and dimeric states: $\mu^\circ_{\mon_2} - 2\mu^\circ_{\mon_1} = 0$ and $2\mu^\circ_{\mona_2} - \mu^\circ_{\mona_1} = 0$.
As a consequence, the energy flow vanishes at any value of the force, and the only source of free energy for the subnetwork $\subx$ is the (always positive) information flow.
In such a case, the subnetwork $\suby$ acts as a Maxwell Demon powered by the chemostats and transducing free-energy towards the subnetwork $\subx$ purely in form of information.
The value of the force $F$ maximizing the concentration of the target species $\mon_2$ turns out to be $67$ J K$^{-1}$mol$^{-1}$.

In the third scenario (Fig.~\ref{fig:drivenSA}d), there still is no energetic preference between the non-activated monomer and non-activated dimer ($\mu^\circ_{\mon_2} - 2\mu^\circ_{\mon_1} = 0$), but the activated dimer is unfavored with respect to the activated monomer ($\mu^\circ_{\mona_2} - 2\mu^\circ_{\mona_1} > 0$).
As a consequence, the energy flow is always negative, and the term $\IF$ must always be positive and larger than $|\EF|$ in order to keep the subnetwork $\subx$ out of equilibrium.
The value of the force $F$ maximizing the concentration of the target species $\mon_2$ turns out to be $68$ J K$^{-1}$mol$^{-1}$.

In all the three scenarios, the internal free-energy transduction mechanism gets stalled for large values of the external force ($F\gtrsim$ 100 J K$^{-1}$mol$^{-1}$):
$\EF$ and $\IF$ go to zero.
Thus, the subnetwork $\subx$ approaches equilibrium ($\eprx \approx 0$), despite the subnetwork $\suby$ being far from equilibrium.
This can also be understood in terms of the so-called negative differential response of the steady state current~\cite{falasco2019ndr}:  $\currs$ decreases when increasing $F$.

In all the three scenarios, the efficiency $\eta$ is finite for small values of the external force, presents a maximum for intermediate values and then vanishes at large values.
As shown in Appendix~\ref{app:linear}, $\eta$ is constant in the linear regime for very small forces.
The fact that the efficiency is maximized for finite values of the force is therefore a truly far-from-equilibrium feature, which in this specific case can be ascribed to the negative differential response of the current $\currs$ with respect to the force.
Interestingly, in the third scenario (Fig.~\ref{fig:drivenSA}d), the value of the force maximizing the efficiency gets close to the one maximizing the concentration of the target species $\mon_2$, indicating the possibility to fine-tune the parameters in order to optimize both the kinetic and the thermodynamic performance.

As a final comment, we notice that in the second and third scenarios the information flow constitutes the only free-energy source keeping the subnetwork $\subx$ out of equilibrium.
This is in line with the fact that the functioning of this model is often described in terms of an information ratchet mechanism, hinting that the generation of information is the key working principle.
However, the first scenario shows that there may be regimes where the information flow is negative or null, meaning that the information flow does not power the assembling reactions, but rather represents a cost.
We therefore conclude that the thermodynamic concept of information flow is not synonymous with the kinetic concept of information ratcheting, but they are related.
In particular, the connection appears to be vague only in those conditions where a positive energy flow (i.e., an energy ratchet like contribution) is present.

\subsection{Light driven directional bimolecular motor}
\label{sub:lightdrivenmotor}

\begin{figure*}[t]
        \includegraphics[width=.98\textwidth]{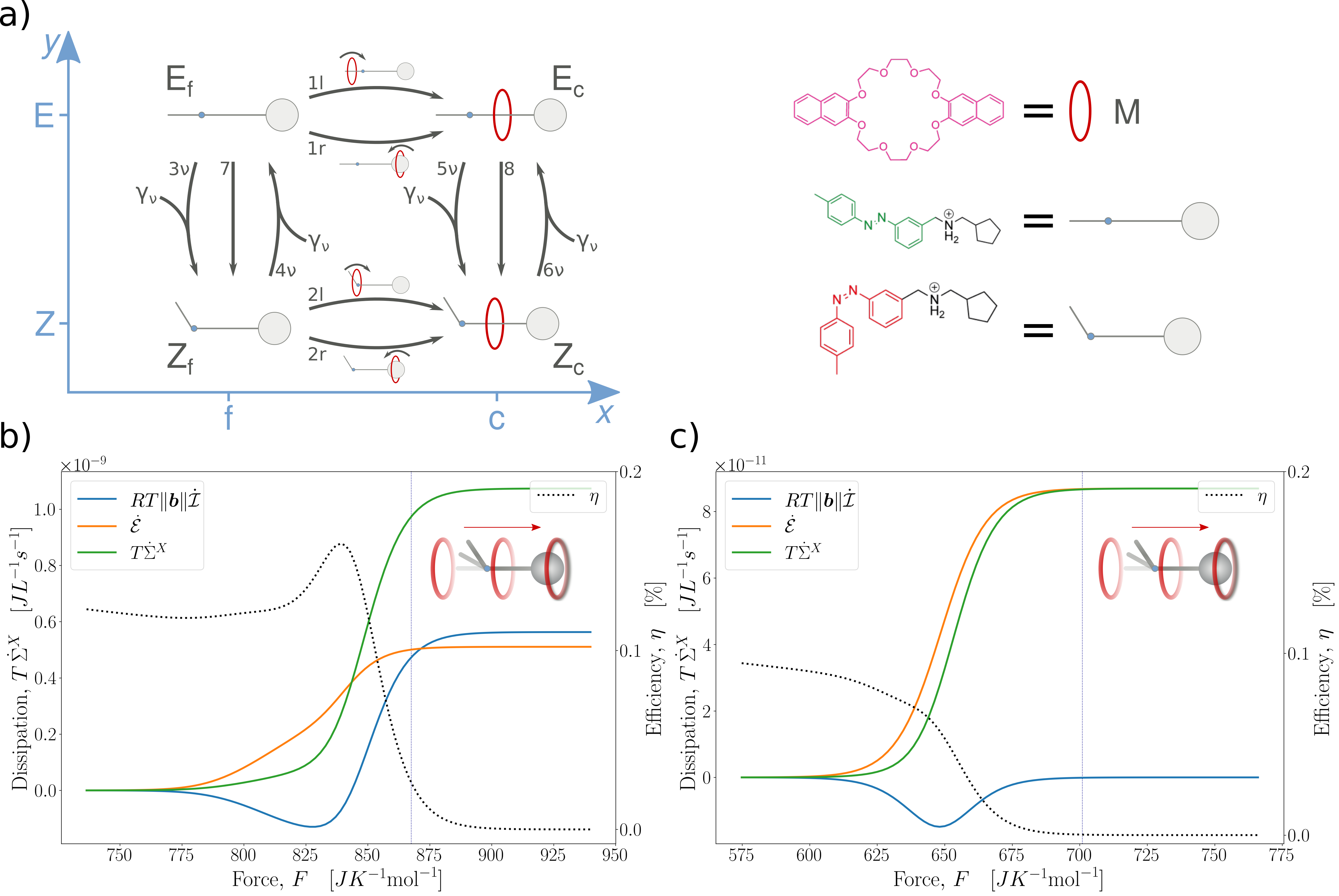}
        \centering
        \caption{
            \textbf{Light-driven bimolecular motor.}
a) Hypergraph representation of the bipartite chemical reaction network and chemical structures of the involved species.
Arrows are used to indicate the conventional direction of the reactions, which are all reversible (photochemical reactions are theoretically reversible~\cite{penocchio2021photo}, but practically irreversible as explained in~\ref{app:photocurr}).
Reactions are labelled according to equations~\eqref{rct:threadings}, \eqref{rct:photoisomerizations}, and~\eqref{rct:thermalisomerization};
b,c) Numerical simulations of the motor operated at 365 nm (b) and 436 nm (c). Information flow ($\IF$), energy flow ($\EF$), and dissipation of the self-assembly reactions ($T\eprx = \EF + \IF$) at steady state are plotted against the net thermodynamic force acting on the system ($F = \mu_\nu / T$, with $T = 298$K) in a range corresponding to a photon flow from $10^{-15}$ to 10$^{-7}$ moles of photons per second.
Experimentally probed conditions are marked by the vertical dotted lines.
The efficiency of the internal free-energy transduction ($\eta$) is computed according to Eq.~\eqref{eq:effphoto}.
Experimental parameters in both the regimes are taken from the original experimental data in Ref.~\cite{ragazzon2015} and Refs.~\cite{sabatino2019,asnicar2022}.
At steady state, on-average directed motion of the macrocycles $\ch{M}$ with respect to the axles is achieved, as a clockwise current arises with respect to the hypergraph and reactions $1r$ and $2l$ are kinetically unfavorable compared to reactions $1l$ and $2r$.
        }
        \label{fig:photomotor}
\end{figure*}

As a second application, we consider a class of synthetic bimolecular motors powered by light~\cite{ragazzon2015,credi2021,corra2021}.
Their functioning consists in the autonomous directional threading and dethreating of a crown-ether macrocycle through an asymmetric molecular axle existing in two different conformations~\cite{sabatino2019,asnicar2022}.

This class of systems can be represented by the CRN in Fig.~\ref{fig:photomotor}a.
The threading of the macrocycle (or ring) $\A$ through the free axle, either in the conformation $\E$ or $\Z$, can be described by four different self-assembly reactions forming the ring-axle complex, either in the conformation $\CE$ or $\CZ$:
\begin{equation}
\begin{split}
    &\schemestart $\A$ + $\E$\arrow{->[][$1l$]}$\CE$\schemestop \, , \quad \schemestart $\A$ + $\E$\arrow{->[][$1r$]}$\CE$\schemestop  \\
    &\schemestart $\A$ + $\Z$\arrow{->[][$2l$]}$\CZ$\schemestop \, , \quad \schemestart $\A$ + $\Z$\arrow{->[][$2r$]}$\CZ$\schemestop \, ,
\end{split}
\label{rct:threadings}
\end{equation}
where labels $l$ and $r$ discriminate over which side of the axle the threading/de-threading of the ring occurs (see Fig.~\ref{fig:photomotor}a).
The reactions~\eqref{rct:threadings} follow mass action kinetics.
The switching of the axle between its conformations in the free state as well as in the complex state can happen via four (in practice irreversible) coarse-grained photochemical reactions (see Appendix~\ref{app:photocurr} for the expression of their currents), all triggered by photons $\gamma_\nu$ at frequency $\nu$:
\begin{equation}
\begin{split}
&\schemestart $\E$\arrow{-U>[$\gamma_\nu$][]}$\Z$\schemestop \,,
\makeatletter
\chemmove{\node[below] at (Uarrow@arctangent) {\footnotesize{$3\nu$}}; }
\makeatother \, \quad 
\schemestart $\Z$\arrow{-U>[$\gamma_\nu$][]}$\E$\schemestop \,,
\makeatletter
\chemmove{\node[below] at (Uarrow@arctangent) {\footnotesize{$4\nu$}}; }
\makeatother 
 \\
\quad \\
&\schemestart $\CE$\arrow{-U>[$\gamma_\nu$][]}$\CZ$\schemestop \,,
\makeatletter
\chemmove{\node[below] at (Uarrow@arctangent) {\footnotesize{$5\nu$}}; }
\makeatother \, \quad 
\schemestart $\CZ$\arrow{-U>[$\gamma_\nu$][]}$\CE$\schemestop  \, .
\makeatletter
\chemmove{\node[below] at (Uarrow@arctangent) {\footnotesize{$6\nu$}}; }
\makeatother
\end{split}
\label{rct:photoisomerizations}
\end{equation}
Changes of the axle's conformation are also possible via thermal reactions following mass-action kinetics:
\begin{align}
    \schemestart $\CE$\arrow{->[][$8$]}$\CZ$\schemestop  \, , \quad \schemestart $\E$\arrow{->[][$7$]}$\Z$\schemestop \, .
\label{rct:thermalisomerization}
\end{align}
Finally, four ``futile''  photochemical reactions are also possible, where photons are absorbed without leading to a conformation change of the axle:
\begin{equation}
\begin{split}
&\schemestart $\E$\arrow{-U>[$\gamma_\nu$][]}$\E$\schemestop 
\makeatletter
\chemmove{\node[below] at (Uarrow@arctangent) {\footnotesize{$9\nu$}}; }
\makeatother \, \quad 
\schemestart $\Z$\arrow{-U>[$\gamma_\nu$][]}$\Z$\schemestop
\makeatletter
\chemmove{\node[below] at (Uarrow@arctangent) {\footnotesize{$10\nu$}}; }
\makeatother \\
&\schemestart $\CE$\arrow{-U>[$\gamma_\nu$][]}$\CE$\schemestop 
\makeatletter
\chemmove{\node[below] at (Uarrow@arctangent) {\footnotesize{$11\nu$}}; }
\makeatother \, \quad 
\schemestart $\CZ$\arrow{-U>[$\gamma_\nu$][]}$\CZ$\schemestop 
\makeatletter
\chemmove{\node[below] at (Uarrow@arctangent) {\footnotesize{$12\nu$}}; }
\makeatother \, .
\end{split}
\end{equation}

For this CRN, we define the set of bipartite species as $\setspeciesb = \{\E, \Z, \CE, \CZ\}$ and the set of ancillary species as $\setspeciesa = \{ \A \}$.
Accordingly, reactions can be split into three sets.
The futile photochemical reactions $\setrcta = \{ 9\nu,10\nu,11\nu,12\nu \}$ do not lead to any change in the concentrations, but they contribute to the dissipation of the CRN, as shown in the following.
The reactions $\setrctx = \{1l, 1r, 2l, 2r \}$ interconvert the assembled state of the bipartite species $\indexx \in \{ \mathrm{f}, \mathrm{c}\}$ between the free and the complex state. 
They thus define the self-assembly subnetwork $\subx$.
The reactions $\setrcty = \{3\nu, 4\nu, 5\nu, 6\nu, 7, 8 \}$ interconvert the isomerization state of the bipartite species $\indexy \in \{ \mathrm{E}, \mathrm{Z}\}$.
They thus define the isomerization subnetwork $\suby$.

As done for the previous example, we study the network's energetics at steady state by specializing Eqs.~\eqref{eq:final_deco_x},~\eqref{eq:final_deco_y} and~\eqref{eq:final_deco_a} 
{\small
\begin{subequations}
\begin{align}
T\eprx & = \underbrace{\currs(\mu^\circ_{\CZ} - \mu^\circ_{\CE} + \mu^\circ_{\E} - \mu^\circ_{\Z})}_{=:\dot{\mathcal{E}}} + \underbrace{RT\currs \ln \frac{[{\E}][{\CZ}]}{[{\Z}] [\CE]}}_{=:RT\nrmt\dot{\mathcal{I}}} \,, \\
T\epry & = \dot{W}^\setrcty_\nu - \dot{\mathcal{E}} - RT\nrmt\dot{\mathcal{I}} \,, \\
T\epra & = \dot{W}^\setrcta_\nu\,,\label{eq:diss_a} 
\end{align}\label{eq:secondlaw_photo}
\end{subequations}}
with $\currs= \currs_{1l} + \currs_{1r} = -(\currs_{2l} + \currs_{2r}) = -\currs_{3\nu} -\currs_7 +\currs_{4\nu} = \currs_{5\nu} +\currs_7 -\currs_{4\nu}$ the steady state net current (considered as positive when flowing clockwise across the hyper-graph in Fig.~\ref{fig:photomotor}a),
$\dot{\mathcal{I}} =- \infoflowx = \infoflowy$, and
$\dot{\mathcal{E}}$ the energy flow.
Furthermore, we introduced the two work sources $\dot{W}^\setrcta_\nu$ and $\dot{W}^\setrcty_\nu$ quantifying the rates at which free-energy is provided by the radiation to futile and photoisomerization reactions, respectively (see Eq.~\eqref{eq:photowork_futile} and~\eqref{eq:photowork} for their expressions). 
The former is fully dissipated by the futile reactions (see Eq.~\eqref{eq:diss_a}) and, consequently, it 
is useless to the motor's functioning.
The latter accounts for the fraction of the free-energy absorbed from the radiation which can actually be used to sustain the motor's functioning~\cite{corra2021}.
Similarly to the previous example, the $\suby$ subnetwork (involving the photoisomerization reactions) is the only one coupled to the external force $F = \mu_\nu / T$, but part of the free-energy exchanged with the radiation can be transduced towards subnetwork $\subx$ (involving self-assembly reactions) through the energy $\EF$ and information $\dot{\mathcal{I}}$ flows.
These two mechanisms are the only possible free-energy sources for subnetwork $\subx$ which can be used at steady state to sustain the self-assembly reactions, i.e., the motor's functioning.

Based on the above understanding, the efficiency at which the free-energy harvested by subnetwork $\suby$ from the radiation is transduced into free-energy available to subnetwork $\subx$ can be defined as:
\begin{align}
   0 \leq \eta = \frac{RT\nrmt\dot{\mathcal{I}} + \dot{\mathcal{E}}}{\dot{W}^\setrcty_\nu} = 1-\frac{T\epry}{\dot{W}^\setrcty_\nu} \leq 1 \, .
   \label{eq:effphoto}
\end{align}
which is analogous to Eq.~\eqref{eq:effsa} except that the source work $\dot{W}^\setrcty_\nu$ is performed by the radiation instead of the chemostats.
Similar comments on the properties of the efficiency $\eta$ as those for the case of driven self-assembly apply here as well.

To illustrate our results, we simulated the CRN in Fig.~\ref{fig:photomotor}a under two previously characterized experimental regimes which just differ by the wavelength ($\lambda$, assumed as monochromatic) of the radiation~\cite{ragazzon2015}.
Based on experimental measures, the threading of the ring through the axles is always energetically favored independently of the isomerization state, but it is more favored in the $\ch{E}$ state, namely $\mu_\CE^\circ - \mu_\E^\circ  - \mu_\A^\circ {<} \mu_\CZ^\circ - \mu_\Z^\circ - \mu_\A^\circ {<} 0$.
Furthermore, because of steric hindrance, threading in the $\ch{E}$ state happens preferentially via reaction $1l$ in Eq.~\eqref{rct:threadings}, and de-threadering in the $\ch{Z}$ state happens preferentially via reaction $2r$.
Such a kinetic preference favors the net directed motion of the ring with respect to the axles when the CRN is brought out of equilibrium.
This is often recognized as an energy ratchet mechanism in the literature of light-driven molecular motors, and provides the kinetic asymmetry allowing the the motor's functioning~\cite{zerbetto2007,ragazzon2015,astumian2016,credi2021,credi2021pumps}.
From a thermodynamic standpoint, this corresponds to a positive energy flow $\EF$ for any value of the external force since i) $(\mu^\circ_{\CZ} - \mu^\circ_{\CE} + \mu^\circ_{\E} - \mu^\circ_{\Z})>0$ and ii) $\currs>0$ (see Eq.~\eqref{eq:secondlaw_photo}).

In the first regime ($\lambda = 365$nm, Fig.~\ref{fig:photomotor}b), the complexes $\CE$ and $\CZ$ are more effective than their free counterparts $\E$ and $\Z$ in absorbing light, thus causing photochemical reactions $5\nu$ and $6\nu$ in Eq.~\eqref{rct:photoisomerizations} to occur faster than the photochemical reactions $3\nu$ and $4\nu$ in Eq.~\eqref{rct:photoisomerizations}.
This is often recognized as an information ratchet mechanism providing an additional kinetic bias to the energy ratchet mechanism~\cite{zerbetto2007,ragazzon2015,astumian2016,credi2021pumps,credi2021}.
In the second regime ($\lambda = 436$nm, Fig.~\ref{fig:photomotor}c), the photochemical properties of the axles are independent of the assembled state and, therefore, the motor's functioning is achieved by virtue of the energy ratchet mechanism alone~\cite{ragazzon2015}.
In both cases, we find that relatively high values of the force are needed to take the self-assembly subnetwork out of equilibrium ($T\eprx >0$) and, in contrast to the previous example, the dissipation increases monotonically up to a plateau, as well as the energy flow.
On the other hand, the information flow is initially negative and reaches a plateau only after displaying a minimum.
The plateau value is positive when the irradiation wavelength is 365 nm, while the information flow is always negative when the irradiation wavelength is 436 nm.
As in the previous example, connections can be drawn between information/energy flow and information/energy ratcheting, but clearly the two concepts do not coincide.
In particular, when the mechanism is a pure energy ratchet ($\lambda$ = 436 nm, Fig.~\ref{fig:photomotor}c), the only positive contribution to $T\eprx$ comes from the energy flow.
When information ratcheting is added ($\lambda$ = 365 nm, Fig.~\ref{fig:photomotor}{\color{black}b}) regimes with positive information flow arise.
However, when both energy and information ratchet mechanisms are active, regimes where the information flow is null or negative are observed, and when there is no information ratchet mechanism, the information flow is always not null and negative.

Finally, in both cases the efficiency stays finite in the linear regime (see Appendix~\ref{app:linear}) and drops to zero at high forces as $T\eprx$ saturates, with values of $3\times10^{-2}\%$ for $\lambda$ = 365 nm and $3\times10^{-4}\%$ for $\lambda$ = 436 nm at the experimentally probed conditions~\cite{ragazzon2015} (marked by the vertical dotted lines in Fig.~\ref{fig:photomotor}b,c).
Interestingly, while at $\lambda$ = 436 nm the efficiency drops to zero monotonically, it presents a maximum at intermediate forces for $\lambda$ = 365 nm.


\section{Conclusion}

In this work, we formulated information thermodynamics for deterministic bipartite CRNs with nonlinear dynamics.
The resulting picture is that these networks can be divided into two subnetworks made of different sets of reactions, which interact (due to the presence of shared species) by exchanging free-energy in the form of information and/or energy flows.
To get there, we defined the mutual information in terms of the concentrations of bipartite species (Eq.~\eqref{eq:info_def}), showing that it can be expressed in terms of Shannon-like contributions to the total entropy (Eq.~\eqref{eq:decomp_shannon_entropy}).
Furthermore, we specialized the second law for each subnetwork (Eq.~\eqref{eq:final_deco}), in analogy with information thermodynamics of Markov jump processes.
Crucially, informational terms appear and may play the role of free-energy sources sustaining reactions which would not occur spontaneously.
At odds with information thermodynamics of stochastic processes, information flow terms in deterministic CRNs must account for non-normalized concentration distributions.
Also, variation of the subnetworks entropy due to the standard molar entropy carried by the chemical species and the contribution of non-bipartite species which may be present have to be taken into account.

Our framework generalizes a previous formulation developed for linear dynamics~\cite{amano2021info}.
It directly applies to most of the artificial molecular machines and motors reported until now, and to many relevant models in systems chemistry and biochemistry too (e.g., biomolecular motors, signal transduction).
Here, we analyzed two epitomes of nonequilibrium supramolecular chemistry where the nonlinear self-assembly steps explicitly require leveraging our new results, as standard methods of information thermodynamics fail to treat non-normalized concentration distributions.
Unprecedented insights were provided on i) how free-energy is used to power the chemically-driven self-assembly of monomers and an experimental light-driven bimolecular motor, and ii) on the corresponding performance evaluated according to their thermodynamic efficiencies that we introduced (Eqs.~\eqref{eq:effsa} and~\eqref{eq:effphoto}).
Furthermore, the role of information as a physical quantity is clarified and compared with the concept of information ratcheting.
Albeit related, we found counterexamples to a general one-to-one correspondence between the two, thus establishing that the two concepts leverage two different notions of information, thus pertaining to distinct {\color{black}levels} of descriptions (a thermodynamic level and a kinetic one).

Our results can be straightforwardly extended to cases where the external parameters $\concc$ and $n_\nu$ change in time according to a time dependent protocol~\cite{forastiere2020}, as well as when fluxes between the system and the reservoirs are controlled instead of the concentrations~\cite{avanzini2022flux}.
{\color{black}Also}, if nonideal solutions were considered~\cite{Avanzini2021}, additional terms would appear in the expression of the subnetworks entropy production without qualitatively altering the results.
{\color{black}Furthermore, our approach should apply to multipartite CRNs too, namely when species and reactions can be split into more than two subnetworks exchanging information (e.g., a bipartite network where the species can also diffuse between two compartments defining an additional label for both species and reactions)~\cite{horowitz2015,wolpert2020}.
However, additional work is required to meaningfully explore this direction, especially concerning the notions of thermodynamic efficiency.}

It is now clear that information processing is a key feature of living systems which is prone to quantitative analyses~\cite{bialek2016,emonet2021}.
With our work, we developed some of the theoretical tools needed to extend the approach of information thermodynamics towards the deterministic domain.
As concentration distributions are often easier to characterize than probability distributions, we foresee that the current understanding of the role of information in chemical systems can benefit to some extent from the results that we derived.

\begin{acknowledgments}
This research was funded by the Luxembourg National Research Fund, grant ChemComplex (C21/MS/16356329), and by the FQXi Foundation, grant ‘Information as a fuel in colloids and superconducting quantum circuits’ (FQXi-IAF19-05).
\end{acknowledgments}

\appendix

\section{Photochemistry}

In this appendix we derive the main quantities needed to reproduce the results in Section~\ref{sub:lightdrivenmotor}, namely, the expressions of photochemical reaction currents and the photochemical work.

\subsection{Photochemical currents}
\label{app:photocurr}

In a typical photochemical experiment of the kind considered in Section~\ref{sub:lightdrivenmotor}, the solution is contained in a square \textit{cuvette} and irradiated from one side with a fixed photon flow $I_0$ quantifying the amount of moles of photons impinging on the system's surface $\Omega$ per unit time~\cite{Montalti2006}.
The radiation frequency is selected through interference filters with a bandwidth $\Delta \nu$ of approximately 10 nm in unit of wavelength~\cite{Montalti2006}, and all the frequency-dependent properties of the system are considered as constants over this interval.
This approximation is equivalent to assuming the radiation to be monochromatic at the chosen frequency.
According to the Beer-Lambert law, the intensity of the photon flow decreases along the direction of propagation $r$ 
(with $r=0$ being the point where the radiation starts interacting with the solution and $r = r_f$ being the optical path) as:
\begin{equation}
    I(r) = I_0 \cdot 10^{-r \boldsymbol{\epsilon} \cdot \boldsymbol{b}} \, ,
    \label{eq:lambertbeer}
\end{equation}
where ${\boldsymbol \epsilon}$ is the vector collecting the molar extinction coefficients of each species.
Note that the concentrations $\boldsymbol b$ are considered {\color{black}uniform in the sample (i.e., we consider a well-mixed solution)}.
To express the current of the photoisomerization reaction $\mathrm{E} \ce{->[\text{light}]} \mathrm{Z}$, we compute the number of photons absorbed per unit of time and volume by the species $\mathrm{E}$, and we multiply it by the quantum yield $\phi_\mathrm{E \rightarrow Z}$ of the process, accounting for the probability that the photoisomerization happens once $\mathrm{E}$ absorbed a photon~\cite{Mauser1998}:
\begin{equation}
    \currs^\mathrm{E \rightarrow Z} = \frac{I_0}{V} (1 - 10^{-r_f\boldsymbol{\epsilon} \cdot \boldsymbol{b}}) \frac{\epsilon_\mathrm{E} [\mathrm{E}]}{\boldsymbol{\epsilon} \cdot \boldsymbol{b}} \phi_\mathrm{E \rightarrow Z}\,.
    \label{eq:photochemicalrate_nonlinear}
\end{equation}
The above expression simplifies under the the working conditions of interest~\cite{ragazzon2015}, i.e., the total absorbance of the solution at steady state is small ($r_f \boldsymbol{\epsilon} \cdot \boldsymbol{b} \ll 1 $), to
\begin{equation}
    \currs^\mathrm{E \rightarrow Z} =  \frac{\ln(10) r_f \epsilon_\mathrm{E} \phi_\mathrm{E \rightarrow Z}}{V} I_0 [\mathrm{E}] = k_\mathrm{E \rightarrow Z} I_0 [\mathrm{E}] \, ,
    \label{eq:photochemicalrate}
\end{equation}
which is the expression we used in our simulations in Sec.~\ref{sub:lightdrivenmotor}.
The currents of other photoisomerization reactions are derived in the same way,
as well as the currents of futile photochemical reactions, like for $\mathrm{E} \ce{->[\text{light}]} \mathrm{E}$
\begin{equation}
    \currs^\mathrm{E \rightarrow E} =  \frac{\ln(10) r_f \epsilon_\mathrm{E} \phi_\mathrm{E \rightarrow E}}{V} I_0 [\mathrm{E}] = k_\mathrm{E \rightarrow E} I_0 [\mathrm{E}] \, ,
    \label{eq:futilerate}
\end{equation}
with $\phi_\mathrm{E \rightarrow E} = 1 -\phi_\mathrm{E \rightarrow Z}$.
Equations~\eqref{eq:photochemicalrate} and~\eqref{eq:futilerate} show that photochemical reactions in low-absorbance regimes under controlled constant monochromatic irradiation can be treated as irreversible unimolecular mass action reactions~\cite{Mauser1998,Montalti2006} by introducing effective kinetic constants, e.g., $k_\mathrm{E \rightarrow Z}$ and $k_\mathrm{E \rightarrow E}$.
We note that a more detailed analysis taking into account the underlying excited state dynamics (photoisomerization reactions usually follow the so-called diabatic mechanism~\cite{Balzani2014}) would show that this kind of reactions are in principle reversible, but the backward pathways (e.g., $\mathrm{Z}$ gets thermally excited and decays towards $\mathrm{E}$ by emitting a photon) are so unlikely that they can be safely neglected for any practical purpose, thus further justifying the above derivation~\cite{penocchio2021photo}.

\subsection{Photochemical work}
\label{app:photowork}
The concentration distribution of photons in the system can be expressed as $n_\nu(r) = n_\nu^T + n_0 10^{-r \boldsymbol{\epsilon} \cdot \boldsymbol{b}}$, where the black body distribution $n_\nu^T$ at the temperature $T$ and $n_0$ account for the contribution due to the thermal bath and the external radiation, respectively.
Within the experimental approximations, the latter can still be expressed in terms of a black body distribution ($n_0 = n_\nu^{T_\mathrm{r}}$) by assigning to the external radiation the temperature $T_\mathrm{r}$ at which a black body would emit the same number of photons in the interval selected by the interference filter.
To find $T_\mathrm{r}$, we therefore impose
\begin{equation}
    I_0 = \Omega c \int_{\nu - \Delta \nu/2}^{\nu + \Delta\nu/2} n_{\nu'}^{T_\mathrm{r}} \mathrm{d}\nu' \approx \Omega c n_{\nu}^{T_\mathrm{r}} \Delta \nu \, ,
    \label{eq:effectiveT}
\end{equation}
with $\Omega$ the irradiated area and $c$ the speed of light.
The approximation of treating the the black body distribution as almost constant over the integration interval $\Delta \nu$ is not necessary, but we mention it because it simplifies the identification of $T_\mathrm{r}$ without qualitatively changing the results.
For the plots shown in Fig.~\ref{fig:photomotor}, Eq.~\eqref{eq:effectiveT} has been solved numerically without employing the mentioned approximation. 
In the low absorbance regime of interest in Ref.~\cite{ragazzon2015}, we have that:
\begin{align}
    n_\nu(r) =& n_\nu^T + n_\nu^{T_\mathrm{r}} 10^{-r\boldsymbol{\epsilon} \cdot \boldsymbol{b}} \nonumber \\
    \approx& n_\nu^T + n_\nu^{T_\mathrm{r}} (1 - \ln (10)r\boldsymbol{\epsilon} \cdot \boldsymbol{b}) \, .
\end{align}
As a consequence, the chemical potential of the radiation~\eqref{eq:mu_nu} will be a function of $r$ too.
By considering $\mu_\nu(r)$ to be constant in the region of space from $r$ to $r + \mathrm{d} r$, the free energy absorbed per unit of time from the radiation in the same region and in the low absorbance regime can be expressed as 
\begin{align}
&\mu_\nu(r) (I_0 (10^{-r\boldsymbol{\epsilon} \cdot \boldsymbol{b}} - 10^{-(r+\mathrm{d}r)\boldsymbol{\epsilon} \cdot \boldsymbol{b}})) \nonumber \\ 
&\approx \mu_\nu(r) I_0 \ln (10) \boldsymbol{\epsilon} \cdot \boldsymbol{b} \, \mathrm{d}r\, .
\end{align}
The free energy harvested from the radiation by the whole system per unit time and volume is thus given by
\begin{equation}
    \dot{W}_\nu =\frac{1}{V} \int_0^{r_f}\mu_\nu(r) I_0 \ln (10) \boldsymbol{\epsilon} \cdot \boldsymbol{b} \, \mathrm{d}r\, ,
\end{equation}
and it can be split into the contributions of $\setrcty$ and $\setrcta$ reactions as
\begin{equation}
    \dot{W}_\nu = \dot{W}_\nu^\setrcty +  \dot{W}_\nu^\setrcta \, ,
\end{equation}
with
\begin{equation}
    \dot{W}_\nu^\setrcty = \sum_{\rho \in \setrcty_{\mathrm{ph}}} \frac{\currs^{\color{black}\rho}}{r_f} \int_0^{r_f} \mu_\nu (r) \mathrm{d} r = \sum_{\rho \in \setrcty_{\mathrm{ph}}} \currs^{\color{black}\rho} \langle \mu_\nu (r) \rangle  \, ,
    \label{eq:photowork}
\end{equation}
and
\begin{equation}
    \dot{W}_\nu^\setrcta = \sum_{\rho \in \setrcta} \frac{\currs^{\color{black}\rho}}{r_f} \int_0^{r_f} \mu_\nu (r) \mathrm{d} r = \sum_{\rho \in \setrcta} \currs^{\color{black}\rho} \langle \mu_\nu (r) \rangle  \, ,
    \label{eq:photowork_futile}
\end{equation}
with $\setrcty_{\mathrm{ph}}=\{3\nu,4\nu,5\nu,6\nu\}\subset\setrcty$ the set of photoisomerization reactions in Eq.~\eqref{rct:photoisomerizations}.
Equations~\eqref{eq:photowork} and~\eqref{eq:photowork_futile} allow us to identify the average chemical potential $\langle \mu_\nu (r) \rangle$ as the photochemical force associated to the radiation, i.e., $F = \langle \mu_\nu (r) \rangle / T $, which is the expression used for the results reported in Fig.~\ref{fig:photomotor}.
We stress that this is a consequence of the low absorbance approximation, which makes the photochemical currents $\currs_\rho$ independent of $r$.
If this approximation did not hold, the currents $\currs_\rho$ would enter the integrals in Eqs.~\eqref{eq:photowork} and~\eqref{eq:photowork_futile} and the average chemical potential would not play the role of the net thermodynamic force~\cite{corra2021}.

\section{Linear regime}
\label{app:linear}

In this appendix, we compute the efficiencies~\eqref{eq:effsa} and~\eqref{eq:effphoto} in the linear regime, i.e., when the external force acting on the CRN is small enough ($F \ll R$) that the concentrations at steady state are well approximated by linear shifts from the equilibrium ones proportional to the external force according to
$[\chemspecies]_\mathrm{ss} = [\chemspecies]_\mathrm{eq}(1 + m_{\chemspecies}F/R)$ with $m_{\chemspecies}$ the coefficient of proportionality.
In these conditions, the chemical potentials~\eqref{eq:chempotz} read $\chempotential_{\setspecies} = \chempotential_{\setspecies}^{\mathrm{eq}} + \boldsymbol{m} T F$ with $\boldsymbol{m} = (\dots,m_\chemspecies,\dots)^\intercal$,
while the current of a generic chemical reaction $\rho$ following mass action kinetics can be expressed at the first order in $F$ as
$\currs^\rct = -\currs_\mathrm{eq}^{+\rho}( \matS_\rho\cdot \boldsymbol{m}) F / R$ (where we used $\currs_\mathrm{eq}^{+\rho} = \currs_\mathrm{eq}^{-\rho}$).


\subsection{Driven self-assembly}
Without loss of generality, we consider $[\F] = [\F]_\mathrm{eq} (1 + F / R)$ and $[\W] = [\W]_\mathrm{eq}$, which is consistent with $(\mu_\F - \mu_\W)/T = F$ in the first order in $F$. 
Under these conditions, $\eta$ as computed with Eq.~\eqref{eq:effsa} is constant up to the second order in $F$.
Indeed, for $T\eprx$ we have
{\small
\begin{align}
T\eprx  =& \currs(\mu_\mathrm{\mon_2} - \mu_\mathrm{\mona_2} + 2\mu_\mathrm{\mona_1} - 2\mu_\mathrm{\mon_1}) = \nonumber \\
        =& \frac{TF^2}{R} \currs^{+2}_\mathrm{eq}(2m_{\mona_1} - m_{\mona_2})(m_\mathrm{\mon_2} - m_\mathrm{\mona_2} + 2m_\mathrm{\mona_1} - 2m_\mathrm{\mon_1})\,,
\end{align}}
where the steady state current has been computed from reaction $2$.
For $I_{\ch{F}}(\mu_\F - \mu_\W)$, we have
{\small
\begin{align}
I_{\ch{F}}(\mu_\F - \mu_\W) =& TF (\currs^{3\F} + 2\currs^{4\F}) = \nonumber \\
                            =& \frac{TF^2}{R} (\currs^{+3\F}_\mathrm{eq}(m_{\mon_1} + 1 - m_{\mona_1}) + 2\currs^{+4\F}_\mathrm{eq} (m_{\mon_2} + 2 -m_{\mona_2})) \, .
\end{align}}
Therefore:
\begin{align}
   \eta =\frac{\currs^{+2}_\mathrm{eq}(2m_{\mona_1} - m_{\mona_2})(m_\mathrm{\mon_2} - m_\mathrm{\mona_2} + 2m_\mathrm{\mona_1} - 2m_\mathrm{\mon_1})}{\currs^{+3\F}_\mathrm{eq}(m_{\mon_1} + 1 - m_{\mona_1}) + 2\currs^{+4\F}_\mathrm{eq} (m_{\mon_2} + 2 - m_{\mona_2})}\,.
   \label{eq:effsa_lin}
\end{align}
After analytically computing the coefficients $\boldsymbol{m}$ by solving the steady state of the linearized dynamics, Eq.~\eqref{eq:effsa_lin} yields efficiencies of $7.5\%$, $3.7\%$, and $3.2\%$ for the linear regimes of scenarios b, c and d, respectively.

\subsection{Light driven bimolecular motor}

In the case of the light driven bimolecular motor, we take $n_\nu(r) = n_\nu^T (1 + F 10^{-r\boldsymbol{\epsilon}\cdot \boldsymbol{b}_\mathrm{eq}}/ R)$.
By doing so, the chemical potential~\eqref{eq:mu_nu} at first order in $F$ reads $\mu_\nu (r) = TF 10^{-r\boldsymbol{\epsilon}\cdot \boldsymbol{b}_\mathrm{eq}}$ provided that $f_\nu \gg n_\nu^\mathrm{T}$, as it is the case at both the frequencies considered in Ref.~\cite{ragazzon2015}.
For the equilibrium concentrations $\boldsymbol{b}_\mathrm{eq}$, the low absorbance approximation does not hold, and therefore the free energy absorbed per unit of time from the radiation in the region of space between $r$ and $r + \mathrm{d} r$ reads
\begin{align}
    & \mu_\nu (r) \left( I_0^\mathrm{T} \frac{F}{R} 10^{-r\boldsymbol{\epsilon}\cdot \boldsymbol{b}_\mathrm{eq}} - I_0^\mathrm{T} \frac{F}{R} 10^{-(r + \mathrm{d}r)\boldsymbol{\epsilon}\cdot \boldsymbol{b}_\mathrm{eq}} \right) \nonumber \\
    =& I_0^\mathrm{T} \frac{T F^2}{R} 10^{-2r\boldsymbol{\epsilon}\cdot \boldsymbol{b}_\mathrm{eq}}\ln(10) \boldsymbol{\epsilon}\cdot \boldsymbol{b}_\mathrm{eq} \, \mathrm{d}r \, ,
    \label{eq:linear_dphotowork}
\end{align}
with $I_0^T = \Omega c \int_{\nu - \Delta \nu/2}^{\nu + \Delta \nu/2} n_{\nu'}^T  \mathrm{d}\nu'$.
The free energy harvested from the radiation by the whole system per unit time and volume is obtained by integrating Eq.~\eqref{eq:linear_dphotowork} over $r_f$,
\begin{align}
    \dot{W}_\nu =&  \frac{I_0^\mathrm{T} T F^2 \boldsymbol{\epsilon}\cdot \boldsymbol{b}_\mathrm{eq}\ln(10)}{VR} \int_0^{r_f} 10^{-2r\boldsymbol{\epsilon}\cdot \boldsymbol{b}_\mathrm{eq}}  \, \mathrm{d}r \nonumber \\
    =& \frac{I_0^\mathrm{T} T F^2}{2VR} (1 - 10^{-2r_f\boldsymbol{\epsilon}\cdot \boldsymbol{b}_\mathrm{eq}}) \, ,
\end{align}
and it can be decomposed in the contributions of $\setrcty$ and $\setrcta$ reactions as
\begin{equation}
    \dot{W}_\nu = \dot{W}_\nu^\setrcty +  \dot{W}_\nu^\setrcta \, ,
\end{equation}
with
{
\begin{equation}
    \begin{split}
  \dot{W}_\nu^\setrcty  =& \dot{W}_\nu \Gamma^\setrcty_\mathrm{eq} =  \\
    =&\dot{W}_\nu \Big(\epsilon_\E [\E]_\mathrm{eq} \phi_\mathrm{\E \rightarrow \Z}  + \epsilon_\Z [\Z]_\mathrm{eq} \phi_\mathrm{\Z \rightarrow \E} + \\
    & + \epsilon_\CE [\CE]_\mathrm{eq} \phi_\mathrm{\CE \rightarrow \CZ} + \epsilon_\CZ [\CZ]_\mathrm{eq} \phi_\mathrm{\CZ \rightarrow \CE} \Big)/ \boldsymbol{\epsilon}\cdot \boldsymbol{b}_\mathrm{eq} 
\end{split}
\label{eq:WY_linear}
\end{equation}
}%
and
{
\begin{equation}
    \begin{split}
  \dot{W}_\nu^\setrcta  =& \dot{W}_\nu \Gamma^\setrcta_\mathrm{eq} =  \\
    =&\dot{W}_\nu \Big(\epsilon_\E [\E]_\mathrm{eq} \phi_\mathrm{\E \rightarrow \E}  + \epsilon_\Z [\Z]_\mathrm{eq} \phi_\mathrm{\Z \rightarrow \Z} + \\
    & + \epsilon_\CE [\CE]_\mathrm{eq} \phi_\mathrm{\CE \rightarrow \CE} + \epsilon_\CZ [\CZ]_\mathrm{eq} \phi_\mathrm{\CZ \rightarrow \CZ} \Big)/ \boldsymbol{\epsilon}\cdot \boldsymbol{b}_\mathrm{eq} \,,
\end{split}
\label{eq:WA_linear}
\end{equation}
}%
where $\Gamma^\setrcty_\mathrm{eq}$ and $\Gamma^\setrcta_\mathrm{eq}$ quantify the fractions of the power which are provided by the radiation to photoisomerizations and futile reactions, respectively (with $\Gamma^\setrcty_\mathrm{eq}$ + $\Gamma^\setrcta_\mathrm{eq} = 1$).
Finally, for the sum of the energy and information flows we have:
{\small
\begin{align}
    &T\eprx = \currs (\mu_\mathrm{Z_c} - \mu_\mathrm{E_c} + \mu_\mathrm{E_f} - \mu_\mathrm{Z_f}) = \nonumber \\
    =& \frac{T F^2}{R} (\currs_\mathrm{eq}^{+2l} + \currs_\mathrm{eq}^{+2r}) (m_\mathrm{Z_c} - m_\mathrm{Z_f} - m_\mathrm{M}) (m_\mathrm{Z_c} - m_\mathrm{E_c} + m_\mathrm{E_f} - m_\mathrm{Z_f}) \,.
\end{align}}
Therefore, Eq.~\eqref{eq:effphoto} in the linear regime boils down to
\begin{align} \footnotesize
    \eta = \frac{2 V  (\currs_\mathrm{eq}^{+2l} + \currs_\mathrm{eq}^{+2r}) (m_\mathrm{Z_c} - m_\mathrm{Z_f} - m_\mathrm{M}) (m_\mathrm{Z_c} - m_\mathrm{E_c} + m_\mathrm{E_f} - m_\mathrm{Z_f})}{I^T_0 (1 - 10^{-2r_f\boldsymbol{\epsilon}\cdot \boldsymbol{b}_\mathrm{eq}}) \Gamma^\setrcty_\mathrm{eq}} \, ,
\end{align}
which verifies that the efficiency is constant up to the second order in $F$.
Here, solving the steady state of the linearized dynamics to compute the coefficients $\boldsymbol{m}$ and then find the actual values of $\eta$ in the two experimental regimes would require to know the rates of the backward photochemical reactions, which cease to be negligible at very small forces.




%
\end{document}